\documentclass{article}

\usepackage[utf8]{inputenc}
\usepackage[T1]{fontenc}
\usepackage{PRIMEarxiv}
\usepackage{amsmath,amsfonts}
\usepackage{graphicx}
\graphicspath{{images/}}
\usepackage{booktabs,tabularx,array,makecell}
\usepackage{nicefrac,microtype}
\usepackage{siunitx}
\usepackage[numbers,sort&compress]{natbib}
\usepackage{subcaption}
\usepackage{placeins}
\usepackage{pdflscape}
\usepackage{hyperref}
\usepackage{doi}
\hypersetup{
  pdftitle={A Trans-Domain Digital Twin for Bio-Aware Control of Climate and Energy in Cattle Fattening Barns Using Single-Episode Optimizer Learning},
  pdfauthor={Mansoorali Amiri},
  pdfsubject={Trans-domain digital twins and bio-aware climate control},
  pdfkeywords={Digital Twin, Model Predictive Control, Livestock Growth Simulation, Edge Intelligence, Single-Episode Learning},
  hidelinks
}
\newcolumntype{Y}{>{\raggedright\arraybackslash}X}
\newcolumntype{P}[1]{>{\raggedright\arraybackslash}p{#1}}
\title{A Trans-Domain Digital Twin for Bio-Aware Control of Climate and Energy in Cattle Fattening Barns Using Single-Episode Optimizer Learning}
\author{\href{https://orcid.org/0009-0002-1894-8089}{\includegraphics[height=1.2em,alt={Open Researcher and Contributor ID (ORCID)}]{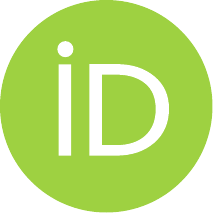}\hspace{1mm}Mansoorali Amiri} \\\
  Montreal, Quebec, Canada}
\date{}

\begin{document}
\maketitle

\begin{abstract}
In closed cattle-fattening barns, the indoor climate and herd growth are interdependent. Temperature, relative humidity, airflow, and ventilation affect thermal comfort, feed intake, metabolic heat production, daily growth, feed efficiency, and energy consumption, while body-weight gain alters the future heat and moisture loads of the barn and, consequently, its ventilation, heating, and energy requirements. This study proposes a trans-domain digital twin framework with single-episode learning capability, designed for bio-aware climate and energy control in a closed cattle-fattening barn. The framework integrates a mechanistic climate simulator, a livestock growth simulator, model predictive control, lightweight reinforcement learning, and structured knowledge storage within a multi-rate control architecture. The fast control loop operates every five minutes to evaluate actuator decisions and maintain short-term thermal comfort, safety, and energy efficiency, whereas the slow control loop provides biological guidance based on daily climatic conditions, feed efficiency, heat production, and growth-limiting factors. The simulation results demonstrate that climate, growth, energy, feed, biological guidance, and memory can be integrated within a single executable control cycle. Further work is required for field validation, improved feed-pressure management, and the reduction of abrupt actuator-command variations.
\end{abstract}


\section{Introduction}

As the global demand for food increases and agricultural labor shortages become a growing constraint, technological innovations are increasingly required to improve the management of complex livestock-production systems~\cite{lee2026agricultural}. In such systems, a digital twin (DT) provides added value when the target process cannot be adequately understood or controlled through simple monitoring or isolated models~\cite{amiri2025jumeaux}.

\subsection{Background and Motivation}

In closed cattle-fattening barns, indoor climate should be understood as part of an integrated climate–biological–energy system rather than as an isolated environmental variable. Temperature, relative humidity, ventilation, and airflow define the physical state of the barn, but their effects extend beyond short-term thermal comfort. In mechanistic livestock production models, climate and genotype act as defining factors of growth capacity, while feed quality and quantity act as limiting factors~\cite{vanderlinden2019ligaps1}. Therefore, barn management cannot rely only on temperature regulation or feed supply; it must consider the interaction among climate, nutrition, physiological response, energy use, and growth performance~\cite{vanderlinden2019ligaps1}.

This interaction becomes critical because environmental temperature, humidity, wind speed, and thermal conditions affect animal comfort, metabolic heat production, maintenance requirements, feed intake, digestion, energy allocation, and average daily gain~\cite{vanderlinden2019ligaps2}. Consequently, a ventilation, heating, or airflow decision may change not only indoor temperature, but also appetite, feed efficiency, heat production, and daily growth~\cite{vanderlinden2019ligaps1, vanderlinden2019ligaps2, vanderlinden2019ligaps3}. At the same time, the biological state of the herd feeds back into the barn climate. As animals grow and gain weight, herd heat production and moisture generation increase, thereby changing thermal load, humidity load, ventilation demand, heating requirement, and future energy consumption~\cite{vanderlinden2019ligaps1, vanderlinden2019ligaps2}. For this reason, a DT for cattle-fattening barns must represent the physical barn state and the biological herd state together, assess how climate decisions affect growth and feed response, and return the effects of growth and heat production to climate management~\cite{vanderlinden2019ligaps1, vanderlinden2019ligaps2, vanderlinden2019ligaps3, amiri2025jumeaux}.

\subsection{Problem Statement}

Despite the effectiveness of model predictive control (MPC) and adaptive control methods in cattle-building climate management, most existing approaches remain limited to regulating temperature, humidity, ventilation, and energy consumption~\cite{drgona2020mpc, yang2019adaptive, xu2024datadriven, bring1999models, tanaskovic2017robust, fiducioso2019safe}. In such approaches, climate is mainly treated as a physical or heating, ventilation, and air conditioning (HVAC)-related control problem, while the biological state of the animal is not explicitly included in the decision-making feedback. However, indoor climate is part of the livestock production system and directly affects thermal comfort, feed intake, digestion, energy allocation, and daily growth~\cite{vanderlinden2019ligaps1, vanderlinden2019ligaps2}.

The main problem is that a climate controller, including an MPC-based controller without growth feedback, may keep the barn temperature within the desired comfort range, but still ignore the effect of the same decision on feed intake, animal heat production, average daily gain (ADG), feed efficiency, and energy/protein limitations~\cite{vanderlinden2019ligaps1,vanderlinden2019ligaps2,vanderlinden2019ligaps3}. In addition, as the herd grows and gains weight, the thermal and moisture loads of the barn change, and future ventilation, heating, and energy requirements are altered~\cite{vanderlinden2019ligaps1, vanderlinden2019ligaps2}. Therefore, a trans-domain digital twin (TDDT) is required to coordinate climate decisions, livestock growth, energy consumption, and feed response within a feedback-based and model-based optimization framework~\cite{vanderlinden2016grass, nguyenky2021indoor, vanderlinden2019ligaps1, vanderlinden2019ligaps2, vanderlinden2019ligaps3, amiri2025jumeaux}.

\subsection{Research Gap}

Despite the broad development of MPC, adaptive control, Bayesian optimization, and building energy models, most existing studies still represent barn climate mainly as a physical–energy control problem. These approaches typically focus on temperature, humidity, ventilation, uncertainty, comfort constraints, and energy cost, while the biological state of livestock is usually not incorporated into the climate decision-making loop~\cite{drgona2020mpc, yang2019adaptive, xu2024datadriven, bring1999models, tanaskovic2017robust, fiducioso2019safe, xin2024review, maddalena2022experimental, lin2023bayesian}. Even in livestock-related climate-control studies, the focus often remains on ventilation, thermal comfort, temperature–humidity index (THI) measures, or indoor air-temperature-based HVAC operation, rather than on the joint effect of climate decisions on animal growth and nutrition~\cite{symeonaki2022ontology, shin2024thi}.

In parallel, mechanistic growth models such as LiGAPS-Beef (Livestock simulator for Generic analysis of Animal Production Systems for beef cattle) can simulate growth, feed intake, digestion, energy and protein use, metabolic heat production, ADG, feed efficiency, and growth-limiting factors~\cite{vanderlinden2019ligaps1, vanderlinden2019ligaps2, vanderlinden2019ligaps3}. However, these models are generally used independently of real-time climate-control decisions. As a result, climate–energy control and livestock growth simulation have usually followed separate paths. The main research gap is therefore the lack of an implementable bio-aware digital-twin architecture that can provide feedback-based coupling among climate, growth, feed, and energy within a unified decision-making framework~\cite{amiri2025jumeaux, subeesh2025agricultural, escriba2024digital}.

\subsection{Objective and Contributions}

The objective of this work is to introduce a Single-Episode Trans-Domain Digital Twin (SE-TDDT) for a closed, 120-head cattle-fattening barn, in which indoor climate, livestock growth, energy consumption, and feed response are coordinated within a unified decision-making framework~\cite{amiri2025jumeaux, kamal2014calving, calo1973growth, maia2005coat, ardicli2018retail}. The proposed architecture is designed to move barn climate control beyond conventional HVAC operation by coupling a lightweight mechanistic climate simulator with a mechanistic livestock growth simulator. The climate simulator represents the short-term barn response to outdoor conditions, actuator commands, and herd thermal load, while the growth layer is based on LiGAPS-Beef and its submodels for BioThermoRegulation, feed intake and digestion, and energy/protein utilization, which provide outputs such as body weight, ADG, feed intake, feed efficiency, heat production, and limiting factors~\cite{nguyenky2021indoor, vanderlinden2019ligaps1, vanderlinden2019ligaps2, vanderlinden2019ligaps3, bring1999models}.

\begin{figure}[htbp]
\centering
\includegraphics[width=0.70\linewidth,alt={Relationship between the fast climate-control loop, daily growth loop, and optimizer.}]{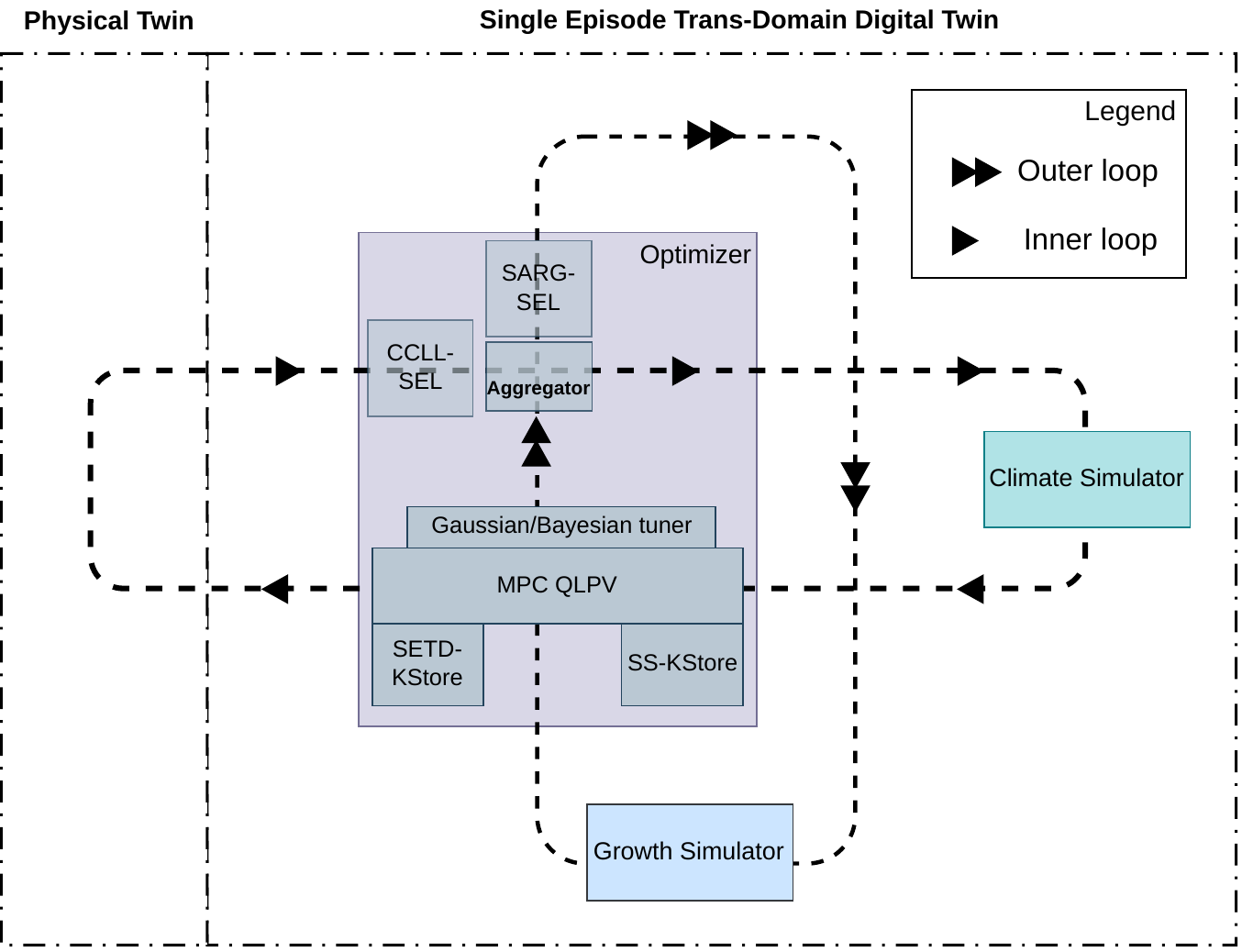}
\caption{Relationship between the inner and outer loops in TDDT optimizer training. QLPV denotes quasi-linear parameter-varying prediction.}
\label{fig:image_1}
\end{figure}

The first contribution is the formulation of a multi-rate SE-TDDT architecture with a fast inner climate-control loop and a slower outer growth-guidance loop. The inner loop evaluates actuator candidates at short time steps to maintain comfort, safety, and energy efficiency, whereas the outer loop aggregates daily climate data and returns biological guidance to the climate optimizer, as illustrated in Figure~\ref{fig:image_1}. The second contribution is the use of MPC as the final actuator-command selection mechanism. MPC selects commands using climate predictions, comfort constraints, energy cost, equipment limits, and biological guidance, while lightweight reinforcement learning (RL) and a Gaussian/Bayesian tuner only adjust decision biases or weights without replacing MPC~\cite{drgona2020mpc, yang2019adaptive, xu2024datadriven, bring1999models, tanaskovic2017robust, fiducioso2019safe}. The third contribution is the integration of single-episode learning with structured knowledge memory. Climate Context Local Library Guided Single-Episode Learning (CCLL-SEL)\footnote{CCLL-SEL is the local climate-context memory. It converts historical and local climate patterns into soft climate priors that help MPC interpret the current barn state under seasonal, humid, warm, cold, or transitional conditions.}, Stage-Aware Reference-Guided Single-Episode Learning (SARG-SEL)\footnote{SARG-SEL is the stage-aware biological guidance component. It uses growth–diet reference paths and livestock stage information to convert growth status, feed response, heat production, and limiting factors into biological guidance for climate-control decisions.}, Step-Stream Knowledge Store (SS-KStore)\footnote{SS-KStore is the streaming memory used during single-episode training. It stores step-wise climate states, actuator decisions, simulator responses, rewards, biological guidance, and daily summaries in a low-memory and append-only form.}, and Single-Episode Trans-Domain Knowledge Store (SETD-KStore)\footnote{SETD-KStore is the final loadable trans-domain knowledge package for online control. It compresses the trained experience into policy snapshots, climate contexts, biological-guidance–decision relationships, reward indicators, coverage, and uncertainty so that the edge controller can start from learned knowledge rather than from scratch.} compress climatic contexts, growth references, training traces, biological guidance, and policy snapshots into a loadable knowledge package for online edge control. In these component names, CCLL denotes Climate Context Local Library, SARG denotes Stage-Aware Reference-Guided, SEL denotes Single-Episode Learning, SS denotes Step-Stream, SETD denotes Single-Episode Trans-Domain, and KStore denotes Knowledge Store. This design allows the edge controller to start from offline SE-TDDT knowledge rather than learning from scratch and is intended to support fine-tuning under real barn conditions.

\section{Related Work}
\subsection{MPC and Adaptive Climate Control}

MPC has been widely used for indoor climate and HVAC control because it can use a mechanistic or learned dynamics model to predict future states of temperature, humidity, thermal load, comfort constraints, and energy consumption, and then select optimal control actions under operational constraints~\cite{drgona2020mpc, bring1999models}. Compared with purely reactive control or unguided exploration methods such as Q-learning over all states, MPC is better suited to managing nonlinear dynamics, thermal inertia, weather variations, uncertainty, internal loads, and equipment constraints~\cite{drgona2020mpc, yang2019adaptive, tanaskovic2017robust, xin2024review}. Complementary learning and tuning methods, including Gaussian process models and Bayesian optimization, have also been used to learn building dynamics or safely tune temperature-control policies ~\cite{xu2024datadriven, fiducioso2019safe, maddalena2022experimental, lin2023bayesian}.

\subsection{Climate Simulation in Livestock Buildings}

The mechanistic climate simulator represents the closed barn as a simplified building–climate–energy system. The simulator follows the logic of single-zone building indoor-climate modeling described by~\cite{bring1999models}, in which the indoor state is determined by heat balance, moisture balance, ventilation and airflow exchange, envelope heat transfer, internal loads, outdoor climate, and actuator operation. Within the current SE-TDDT framework, this simulator estimates the short-term response of indoor temperature, relative humidity, airflow, air quality, thermal load, and controllable energy consumption to candidate actuator commands.

For application in livestock buildings, the simulator is also guided by the barn-calibration logic of~\cite{nguyenky2021indoor}, in which monitored barn data are used to calibrate indoor climate and energy behavior under natural or mechanical ventilation. Accordingly, the model is adapted to barn conditions by accounting for outdoor temperature, humidity, wind, radiation, ventilation pathways, building characteristics, actuator states, and herd heat and moisture loads. In SE-TDDT, this mechanistic simulator has two roles: it serves as the climate prediction model within MPC, and during offline training, it acts as a surrogate physical environment that generates the next simulated barn-climate state after each actuator decision.

Climate simulation in livestock buildings is used to represent the coupled dynamics of temperature, humidity, ventilation, airflow, thermal load, energy consumption, outdoor conditions, and actuator response~\cite{bring1999models, nguyenky2021indoor}. Building climate and energy models can therefore support the evaluation of ventilation strategies, electricity use, and temperature–humidity behavior in livestock barns ~\cite{shin2024thi, costantino2023livestock}.

\subsection{Mechanistic Beef Growth Models}

Mechanistic beef-growth models provide the biological layer required to interpret how barn climate affects livestock performance. In this work, LiGAPS-Beef is used as the livestock-growth domain because it is a climate-based mechanistic and dynamic model that represents potential and feed-limited production at the animal and herd levels, integrating BioThermoRegulation, feed intake and digestion, and energy/protein utilization~\cite{vanderlinden2019ligaps1, vanderlinden2019ligaps2, vanderlinden2019ligaps3}. It produces biological outputs including body weight (BW), ADG, feed intake (FI), feed efficiency (FE), heat production (HP), and growth-limiting factors~\cite{vanderlinden2019ligaps1, vanderlinden2019ligaps2, vanderlinden2019ligaps3}. These outputs explain how changes in weight, heat production, feed response, or biological limitations alter the future climatic requirements of the barn. Therefore, LiGAPS-Beef provides the biological feedback needed to connect growth, feed, and heat production to climate-control decisions~\cite{vanderlinden2019ligaps1, vanderlinden2019ligaps2, vanderlinden2019ligaps3}.

\subsection{Digital Twins in Agriculture}

Digital twins (DTs) in smart agriculture are used to connect monitoring, simulation, prediction, and optimization in decision-oriented systems~\cite{amiri2025jumeaux, subeesh2025agricultural, escriba2024digital}. In controlled agricultural environments, DTs can provide real-time virtual representations of physical and biological processes and support management decisions related to climate, irrigation, nutrition, energy use, and sustainability~\cite{subeesh2025agricultural, escriba2024digital, gonzalez2022monitoring}. However, many existing agricultural DT applications remain domain-oriented and mainly emphasize monitoring or prediction rather than closed-loop coordination among interacting domains. In livestock buildings, this limitation becomes critical because climate, growth, feed response, metabolic heat production, and energy demand are mutually dependent. Therefore, the main limitation of current DT approaches is the weak feedback coupling among climate, growth, feed, and energy within an executable control architecture~\cite{vanderlinden2019ligaps1, vanderlinden2019ligaps2, amiri2025jumeaux, subeesh2025agricultural, escriba2024digital}.

\section{Proposed SE-TDDT Architecture}

\subsection{Overall Architecture}

A DT is a decision-oriented virtual representation of a physical system that links sensing, simulation, prediction, and optimization when the target process cannot be adequately understood or controlled by monitoring alone or by isolated models~\cite{amiri2025jumeaux}. In this study, the Physical Twin (PT) is the closed cattle-fattening barn, including the animals, sensors, actuators, building structure, outdoor weather, and herd context. The DT is the computational counterpart of this barn: it predicts indoor climate, evaluates livestock-growth response, optimizes actuator decisions, and stores operational knowledge for later control.

The architectural foundation of the present work is the TDDT conceptual architecture introduced by Amiri~\cite{amiri2026transdomaindigitaltwinconceptual}. This architecture provides a general framework for coupling heterogeneous domain-specific digital twins through coordinated multi-rate temporal interactions and feedback mechanisms. In the present study, this general TDDT architecture is specialized for the cattle-fattening domain by coupling indoor climate, livestock growth, feed response, metabolic heat production, actuator operation, and energy use. The resulting SE-TDDT constitutes an executable, application-specific realization of the broader TDDT architecture.

A TDDT extends a conventional DT by connecting several heterogeneous but interacting domains within one feedback-based architecture. In the cattle-fattening barn, the relevant domains are not limited to temperature and energy. They include indoor climate, livestock growth, feed response, metabolic heat production, actuator operation, and energy use. Therefore, the TDDT represents the barn as a coupled climate–biological–energy system rather than as a single-domain HVAC process~\cite{amiri2025jumeaux}.

The SE-TDDT is the proposed executable form of this TDDT. It uses one long offline training episode to connect the fast climate-control loop with the slower biological-growth loop, then compresses the learned trans-domain experience into structured memory for online edge control. In this architecture, MPC remains the final actuator-command selection mechanism, while RL, QLPV prediction, CCLL-SEL, SARG-SEL, SS-KStore, and SETD-KStore provide prediction support, biological guidance, lightweight adaptation, and reusable memory. Figure~\ref{fig:image_1} illustrates the relationship between the physical twin, the inner loop, the outer loop, and the SE-TDDT optimizer.

The condensed SE-TDDT architecture is organized into five operational layers. The barn/physical layer provides sensor data, actuator status, outdoor weather, building information, and herd context. The climate simulation layer predicts short-term temperature, humidity, airflow, air quality, energy use, and comfort violations. The growth simulation layer receives daily aggregated climate data and estimates body weight, feed intake, feed efficiency, heat production, growth status, and limiting factors. The decision/optimization layer uses MPC as the final actuator-command selection mechanism, while lightweight reinforcement learning and tuning components only adjust decision biases. The knowledge/memory layer stores climatic contexts, biological guidance, training traces, and policy snapshots through CCLL-SEL, SARG-SEL, SS-KStore, and SETD-KStore.

The overall coupling can be summarized by three compact relations. First, the climate state evolves as

\begin{equation}
x_{t+1}=F_c(x_t,u_t,w_t^{\mathrm{out}},h_d),
\label{eq:climate_transition}
\end{equation}

where $x_t$ is the current state, $u_t$ is the actuator command, $w_t^{\mathrm{out}}$ denotes outdoor weather, $h_d$ is the herd context, and $F_c$ is the climate-transition function. This equation predicts the next barn-climate state after applying an actuator command under outdoor weather and herd load. Second, MPC selects the control action as

\begin{equation}
u_t^*=\mathrm{MPC}(x_t,g_d,K),
\label{eq:mpc_decision}
\end{equation}

where $u_t^*$ is the selected command, $g_d$ denotes the growth state, and $K$ denotes the knowledge package. This equation shows that MPC selects the final actuator command using climate state, herd growth state, and stored knowledge. Third, fast climate records are aggregated daily and transferred to the growth simulator:

\begin{equation}
\bar{x}_d=A(x_t), \quad t \in d.
\label{eq:daily_aggregation}
\end{equation}

Here, $\bar{x}_d$ denotes the daily climate vector and $A(\cdot)$ is the aggregation function. This equation converts five-minute climate records into one daily climate input for the growth simulator. This daily bridge connects the inner climate loop with the outer biological loop. Therefore, as shown in Figure~\ref{fig:image_1}, SE-TDDT transforms barn control from a single-domain HVAC process into a feedback-based trans-domain architecture in which climate decisions, growth response, feed-related guidance, energy use, and memory are coordinated within one executable control cycle.

\subsection{Inner Climate-Control Loop}

The inner loop is the fast climate-control layer of SE-TDDT. It operates at a five-minute time step and converts the current barn condition into an executable actuator command. At each step, it receives indoor temperature, relative humidity, airflow, air quality, previous actuator status, outdoor weather, and herd thermal context. The climate simulator, or its lightweight QLPV surrogate, predicts the short-term effect of candidate actuator commands such as ventilation, heating, dampers, fans, and lighting on temperature, humidity, air quality, energy use, and possible comfort violation~\cite{tanaskovic2017robust}.

MPC then compares the candidate commands using comfort, energy, safety, equipment, and biological-priority terms. The key difference between this inner loop and a reactive HVAC controller is that the decision is not applied immediately after measuring temperature or humidity. Instead, the consequence of each candidate command is first evaluated inside the climate twin, and the lower-risk and lower-cost command is selected for the barn. Biological guidance from the outer loop enters this decision only as a soft correction; MPC remains the final actuator-command selection mechanism.

For each control step, MPC first constructs a limited set of actuator candidates,

\begin{equation}
U_t=\{u_t^{(1)},u_t^{(2)},\ldots,u_t^{(m)}\},
\label{eq:control_candidates}
\end{equation}

where $U_t$ denotes the set of candidate actuator commands, $u_t^{(j)}$ is the $j$th candidate command, $t$ is the control time step, and $m$ is the number of candidates. This equation defines the limited actuator-command candidates evaluated by MPC at each control step. For each candidate, the climate simulator performs a short-term rollout over the MPC horizon,

\begin{equation}
\hat{x}_{t:t+H}^{(j)}=M_c(x_t,u_t^{(j)},w_t,h_d),
\label{eq:climate_prediction}
\end{equation}

where $\hat{x}_{t:t+H}^{(j)}$ denotes the predicted climate trajectory, $M_c$ is the climate simulator, $w_t$ represents outdoor weather, and $H$ is the prediction horizon, measured in control steps. This equation rolls out the predicted climate response of each actuator candidate before real execution. Each candidate is then evaluated by a compact cost function,

To reduce the computational cost of repeatedly calling the mechanistic climate simulator for every MPC actuator candidate, the short-term rollout can be approximated by a QLPV prediction model. In this role, QLPV acts as a lightweight, context-dependent surrogate inside MPC: its coefficients vary with climatic context, herd thermal load, and growth phase, allowing the controller to preserve sensitivity to climatic and biological conditions while remaining suitable for edge execution~\cite{tanaskovic2017robust}.

\begin{equation}
J(u_t^{(j)})=
\alpha_c C_{\text{comfort}}^{(j)}
+\alpha_e C_{\text{energy}}^{(j)}
+\alpha_s C_{\text{safety}}^{(j)}
+\alpha_f C_{\text{conflict}}^{(j)}
-\beta_b B_{\text{bio}}^{(j)} .
\label{eq:mpc_objective}
\end{equation}

Here, $J$ denotes the decision cost, $C_{\text{comfort}}$ the comfort penalty, $C_{\text{energy}}$ the energy cost, $C_{\text{safety}}$ the safety penalty, $C_{\text{conflict}}$ the actuator-conflict term, and $B_{\text{bio}}$ the biological benefit; $\alpha_c,\alpha_e,\alpha_s,\alpha_f,\beta_b$ are the decision weights. This equation scores each actuator candidate by balancing comfort, energy, safety, conflict, and biological benefit. The final actuator command is selected as

\begin{equation}
u_t^*=\arg\min_{u_t^{(j)}\in U_t}J(u_t^{(j)}).
\label{eq:optimal_control}
\end{equation}

This equation selects the actuator command with the lowest total bio-aware MPC cost. Thus, MPC acts as the final decision-maker by selecting the candidate with the lowest combined comfort, energy, safety, and conflict cost after considering biological guidance. The climate simulator has a dual role in this loop: it predicts the short-term consequences of actuator candidates before command execution, and during offline training it also acts as a substitute barn environment that generates the next simulated climate state. This dual use allows SE-TDDT to evaluate actuator decisions safely before deployment while preserving a mechanistic representation of barn climate dynamics.

\subsection{Outer Growth-Guidance Loop}

The outer loop is the slower biological-guidance layer of SE-TDDT. It operates at the daily scale because livestock growth, feed intake, feed efficiency, heat production, and limiting factors become meaningful over accumulated daily conditions rather than at every five-minute climate step~\cite{vanderlinden2019ligaps1, vanderlinden2019ligaps2, vanderlinden2019ligaps3}. Its role is to observe the biological consequences of climate-control decisions and return those consequences to the climate-control logic.

At the end of each day, the high-frequency climate records generated by the inner loop are aggregated into a daily climate vector. This daily vector may include minimum and maximum temperature, mean humidity, mean airflow, ventilation index, energy use, time outside the comfort range, and stress indicators:

\begin{equation}
\bar{x}_d = \Phi_{\text{day}}(X_d,U_d,H_d),
\label{eq:daily_aggregation_inputs}
\end{equation}

where $X_d$ denotes daily climate records, $U_d$ the actuator history, $H_d$ the herd data, and $\Phi_{\text{day}}$ the daily aggregation function. This equation aggregates daily climate records, actuator history, and herd context into the daily growth input. 

The growth simulator receives this aggregated climate information together with diet, herd status, breed assumptions, and growth stage, and computes biological outputs such as body weight, ADG, feed intake, feed efficiency, heat production, and limiting factors~\cite{vanderlinden2019ligaps1, vanderlinden2019ligaps2, vanderlinden2019ligaps3}:

\begin{equation}
Y_d^{\mathrm{growth}} = [\mathrm{BW}_d,\mathrm{ADG}_d,\mathrm{FI}_d,\mathrm{FE}_d,\mathrm{HP}_d,L_d],
\label{eq:growth_output}
\end{equation}

where $\mathrm{BW}_d$ denotes body weight, $\mathrm{ADG}_d$ average daily gain, $\mathrm{FI}_d$ feed intake, $\mathrm{FE}_d$ feed efficiency, $\mathrm{HP}_d$ heat production, and $L_d$ the growth-limiting factors. This equation groups the main daily biological outputs produced by the growth simulator. These outputs are not used only for reporting; they also explain how herd growth and metabolic heat change future barn thermal load and ventilation demand.

Finally, the biological outputs are converted by SARG-SEL and the growth–diet reference paths into biological guidance for the next MPC decisions:

\begin{equation}
\Gamma_d =
[T^{\mathrm{low}}_{\mathrm{pref,d}},T^{\mathrm{high}}_{\mathrm{pref,d}},
\lambda^{\mathrm{comfort}}_d,\lambda^{\mathrm{energy}}_d,
\lambda^{\mathrm{vent}}_d,\beta^{\mathrm{bio}}_d],
\label{eq:guidance_package}
\end{equation}

where $\Gamma_d$ is the guidance package, $T^{\mathrm{low}}_{\mathrm{pref,d}}$ and $T^{\mathrm{high}}_{\mathrm{pref,d}}$ are the comfort bounds, $\lambda^{\mathrm{comfort}}_d$ is the comfort weight, $\lambda^{\mathrm{energy}}_d$ is the energy weight, $\lambda^{\mathrm{vent}}_d$ is the ventilation priority, and $\beta^{\mathrm{bio}}_d$ is the biological bias.  This guidance can adjust the preferred comfort range, comfort weight, energy weight, ventilation or heating priority, and biological bias used by MPC. Therefore, the outer loop acts as a biological translator: it converts climate history into growth indicators and returns growth indicators to the actuator-decision process.

The two loops are therefore complementary. The inner loop answers the short-term control question: which actuator command should be applied now to maintain comfort, safety, and energy efficiency? The outer loop answers the biological question: how did the accumulated climate conditions affect growth, feed response, heat production, and future barn load? Their coupling is the core of SE-TDDT. The five-minute loop generates climate-control experience, the daily loop converts this experience into biological feedback, and the knowledge layer stores the resulting climate–growth–decision relationships for online edge deployment.

\subsection{Multi-Rate Coupling Between Loops}

SE-TDDT uses a multi-rate coupling because barn climate changes at the minute scale, whereas growth, feed intake, feed efficiency, heat production, and limiting factors become meaningful at the daily scale~\cite{vanderlinden2019ligaps1,vanderlinden2019ligaps2,vanderlinden2019ligaps3,amiri2025jumeaux}. Therefore, the five-minute inner loop is connected to the daily outer loop through an aggregation operator that converts high-frequency climate and actuator traces into a daily biological input. The examined coupling path in the verification stage is:

\[
\begin{gathered}
\text{climate simulator} \rightarrow \text{MPC} \rightarrow \text{daily aggregation} \\
\rightarrow \text{growth simulator} \rightarrow \text{SARG-SEL} \rightarrow \text{MPC}.
\end{gathered}
\]

The daily aggregation stage is defined as

\begin{equation}
\bar{x}_d=A(D_d^{\mathrm{in}}), \qquad
D_d^{\mathrm{in}}=\{(x_t,u_t,C_t,E_t)\mid t\in d\},
\label{eq:daily_trace}
\end{equation}

where $A$ denotes the aggregation operator, $D_d^{\mathrm{in}}$ the daily trace, $C_t$ the comfort status, and $E_t$ the energy use. The aggregated vector may include minimum and maximum indoor temperature, mean relative humidity, mean airflow, ventilation index, daily energy use, comfort violation, and stress status. This equation compresses one day of climate, actuator, comfort, and energy traces into a daily biological input. It can then be mapped to the climate format required by the LiGAPS-Beef simulator~\cite{vanderlinden2019ligaps1, vanderlinden2019ligaps2, vanderlinden2019ligaps3}.

\begin{figure}[htbp]
\centering
\includegraphics[width=0.95\linewidth,alt={Three-dimensional scatter plot of twelve climate clusters by temperature, relative humidity, and wind speed, with centroids.}]{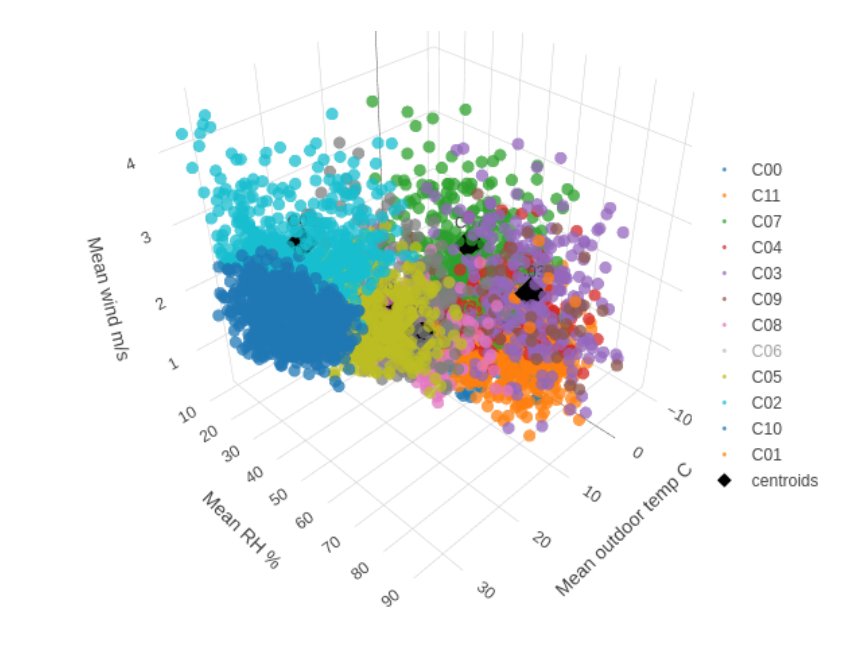}
\caption{CCLL clustering of 12 climatic contexts reconstructed from the 11-year local climate dataset.}
\label{fig:ccll_clusters}
\end{figure}

The growth simulator receives the daily climate vector, herd state, and diet input:

\begin{equation}
y_d^{\mathrm{growth}}=F_{\mathrm{growth}}(\bar{x}_d,g_d,r_d;\Theta_G),
\label{eq:growth_prediction}
\end{equation}

where $y_d^{\mathrm{growth}}$ denotes the growth output, $F_{\mathrm{growth}}$ the growth simulator, $g_d$ the herd state, $r_d$ the diet input, and $\Theta_G$ the growth parameters. This equation updates the daily herd-growth response from aggregated climate, herd state, diet, and growth parameters. Its outputs, including $\mathrm{BW}$, $\mathrm{ADG}$, $\mathrm{FI}$, $\mathrm{FE}$, $\mathrm{HP}$, and limiting factors, are converted by SARG-SEL into biological guidance:

\begin{equation}
b_d=\Phi_{\mathrm{SARG}}(y_d^{\mathrm{growth}},s_d,c_d,K_{\mathrm{SARG}}),
\label{eq:sarg_guidance}
\end{equation}

where $b_d$ denotes biological guidance, $\Phi_{\mathrm{SARG}}$ the guidance mapping, $s_d$ the biological context, $c_d$ the climate context, and $K_{\mathrm{SARG}}$ the reference memory. This guidance returns to MPC as a soft correction of comfort bounds, energy weight, ventilation/heating priorities, and biological bias. Verification of this pathway used five-minute climate traces, daily growth outputs, MPC records, guidance records, SARG-SEL contexts, CCLL-SEL contexts, and KStore artifacts, confirming that the two temporal rates, data ordering, climate/growth units, and guidance path are traceable in the reports. Figure~\ref{fig:ccll_clusters} further supports the coupling between the outer loop and climatic contextual memory through reconstructed CCLL-SEL contexts.

In SE-TDDT, the shared trans-domain data are not only metadata but a trans-domain semantic state that links time, unit, source, quality, climate context, herd state, heat production, feed response, feed efficiency, and biological guidance.

During single-episode training, this semantic state is converted into streaming experience and then stored through SS-KStore and SETD-KStore as reusable climate--growth--decision knowledge for MPC.

The CCLL-SEL contexts shown in Figure~\ref{fig:ccll_clusters} summarize the 11-year local climate memory into compact climate clusters. Each context represents a typical combination of outdoor temperature, relative humidity (RH), and wind speed, and each daily growth step is mapped to the nearest context so that MPC can interpret the current day with respect to historical cold, mild, warm, hot, dry, humid, or windy patterns. In the original CCLL report, the most frequent contexts during the 1000-day trajectory were C10, C00, C05, C11, and C02  in the arid and semiarid climates of the Iranian region; these contexts provide soft priors for ventilation/heating bias and comfort-weight adjustment (Table~\ref{tab:ccll_context_summary}).

SARG-SEL complements the climatic memory of CCLL-SEL by organizing the 1000-day Holstein fattening trajectory into stage-aware biological and diet-guidance phases. While CCLL-SEL maps each day to a climatic context, SARG-SEL maps each day to a growth--feed phase so that the outer loop can interpret whether the herd is in early growth, active gain, high feed and heat-pressure, declining efficiency, or late extended fattening. The SARG-related reports are summarized by the phase-based feed--growth radar plot and the feed--growth--production surface, which show how feed intake, feed efficiency, ADG, heat production, beef-production proxy, heat stress, and feed pressure vary across phases P1--P5. These phase labels are used as soft biological priors for MPC through comfort-range correction, energy weighting, ventilation/heating priority, and biological bias (Table~\ref{tab:sarg_phase_summary}).

\begin{table}[t]
\centering
\small
\caption{Compact SARG-SEL stage--diet interpretation used for biological guidance.}
\label{tab:sarg_phase_summary}
\begin{tabularx}{\linewidth}{@{}P{0.85cm} P{2.4cm} Y Y Y@{}}
\toprule
Phase & Biological stage & Feed--growth interpretation & Main SARG signal & Guidance effect on MPC \\
\midrule
P1 & Initial growth & Early adaptation with increasing body weight and low-to-moderate feed demand. & Growth establishment and feed adaptation. & Keep comfort stable; avoid unnecessary energy use and abrupt ventilation/heating changes. \\

P2 & Active growth & Stronger BW gain with increasing feed intake and improving production response. & Active gain and rising metabolic load. & Maintain comfort while preparing for higher ventilation and heat-load sensitivity. \\

P3 & High growth pressure & High feed pressure and heat production; growth remains active but thermal load increases. & High HP, high FI, and higher biological pressure. & Increase comfort priority and ventilation bias; avoid heat-stress amplification. \\

P4 & Declining efficiency & Growth continues, but ADG and feed efficiency begin to decline gradually. & Reduced FE and weaker gain per feed. & Penalize feed pressure more strongly and balance energy use against growth benefit. \\

P5 & Late extended fattening & BW continues to increase, but ADG and FE are low; feed per kg gain becomes unfavorable. & Late-stage feed inefficiency and high body mass. & Treat decisions as extended-horizon control; reduce feed-pressure risk and smooth actuator changes. \\
\bottomrule
\end{tabularx}
\end{table}

In the compact SARG-SEL graphical abstract, the radar chart compares feed intake, feed efficiency, average daily weight gain, total body-weight gain, beef production index, heat production, heat stress, and feed required per kilogram of weight gain across growth phases P1–P5 (Figure~\ref{fig:integrated_results_e}). The surface plot illustrates how feed pressure, growth efficiency, and the production trajectory are interlinked within the outer loop, and how this relationship supports SARG-SEL biological guidance (Figure~\ref{fig:integrated_results_f}).

\begin{table}[t]
\centering
\small
\caption{Compact summary of CCLL-SEL climatic contexts used as climate priors for MPC.}
\label{tab:ccll_context_summary}
\begin{tabularx}{\linewidth}{@{}P{1cm} P{3cm} P{1.4cm} P{1.4cm} P{1.4cm} Y@{}}
\toprule
Context & Climate type & Mean T (\unit{\degreeCelsius}) & Mean RH (\%) & Mean wind (m/s) & Compact interpretation for MPC \\
\midrule
C00 & Mild / normal humidity / normal wind & 5.420 & 56.588 & 1.118 & Moderately cold context; frequent in the trajectory; useful for stable comfort control. \\
C01 & Cold / normal humidity / normal wind & -0.469 & 64.430 & 1.645 & Cold context; supports heating bias and lower comfort-bound protection. \\
C02 & Hot / dry / normal wind & 25.276 & 34.142 & 2.941 & Hot-dry and windier context; important for ventilation bias and heat-stress control. \\
C03 & Mild / humid / normal wind & 6.701 & 74.972 & 2.323 & Mild-humid context; useful for humidity control and minimum ventilation. \\
C04 & Cold / normal humidity / normal wind & 0.012 & 58.907 & 1.175 & Near-zero cold context; important for heating and cold-bound protection. \\
C05 & Warm / normal humidity / normal wind & 17.523 & 40.568 & 1.493 & Mildly warm and relatively dry context; frequent during growth execution. \\
C06 & Warm / normal humidity / normal wind & 17.912 & 44.883 & 1.889 & Similar to C05 but slightly more humid and windy; separates medium-warm states. \\
C07 & Mild / normal humidity / normal wind & 7.746 & 56.677 & 2.645 & Mild and windier context; affects natural ventilation and heat-loss interpretation. \\
C08 & Warm / normal humidity / normal wind & 15.207 & 53.651 & 1.559 & Mild-warm transitional context between moderate and warm conditions. \\
C09 & Cold / normal humidity / normal wind & 4.406 & 64.369 & 1.638 & Cool-humid normal-wind context; useful for moderate heating and ventilation balance. \\
C10 & Hot / dry / normal wind & 26.756 & 30.441 & 1.741 & Hot-dry context; most frequent in execution; important for ventilation and heat-load bias. \\
C11 & Mild / normal humidity / normal wind & 5.852 & 67.697 & 1.248 & Mild but more humid context; frequent in execution; supports comfort and humidity priors. \\
\bottomrule
\end{tabularx}
\end{table}

\section{Formalism and Decision Logic}

\subsection{Climate State and Transition Model}

At each five-minute control step, the inner loop represents the barn condition by a compact climate–control state vector. This state combines indoor climate, air movement, air quality, previous actuator status, outdoor weather, and herd context:

\begin{equation}
x_t =
\left(
T_t^{\mathrm{in}},
\mathrm{RH}_t^{\mathrm{in}},
v_t^{\mathrm{in}},
q_t^{\mathrm{air}},
k_t,
u_{t-1},
w_t^{\mathrm{out}},
h_d
\right).
\label{eq:climate_state}
\end{equation}

This vector defines the compact barn-climate state observed by MPC at each five-minute step.

Here, $T_t^{\mathrm{in}}$ denotes indoor temperature, $\mathrm{RH}_t^{\mathrm{in}}$ indoor relative humidity, $v_t^{\mathrm{in}}$ indoor airflow, $q_t^{\mathrm{air}}$ air quality, $k_t$ the climate context, and $u_{t-1}$ the previous command.

The short-term climate transition (Eq.~\eqref{eq:climate_transition}) is then written as

\begin{equation}
x_{t+1}=F_c(x_t,u_t,w_t^{\mathrm{out}},h_d).
\label{eq:climate_transition_detail}
\end{equation}

This relation predicts the next climate-control state.

In this relation, $x_{t+1}$ denotes the next state and $F_c$ the climate-transition function. This transition function is the mechanistic climate model used by the inner loop to estimate how actuator decisions, outdoor conditions, and herd thermal load affect the next barn climate state. Therefore, the model provides the predictive basis for MPC while preserving the coupling between physical barn dynamics and biological herd load.

\subsection{Daily Aggregation and Growth Update}

The inner climate loop generates high-frequency climate and actuator records, but the growth simulator operates at the daily scale. Therefore, the five-minute climate states are first compressed into a daily climate vector:

\begin{equation}
\bar{x}_d = A(x_t), \qquad t \in d .
\label{eq:daily_aggregation_detail}
\end{equation}

This aggregation converts high-frequency climate observations into daily inputs.

Here, $t$ denotes the time step and $d$ the day index. The aggregated vector may include daily minimum and maximum temperature, mean relative humidity, mean airflow, ventilation status, comfort violation, and energy use. This vector is the temporal bridge between the climate-control loop and the biological growth loop.

The daily climate vector is then used by the growth simulator to update the livestock biological state:

\begin{equation}
g_{d+1}=F_g(g_d,\bar{x}_d,r_d).
\label{eq:growth_update}
\end{equation}

This relation updates the herd growth state on a daily basis.

In this relation, $g_{d+1}$ denotes the next growth state, $F_g$ the growth simulator, $g_d$ the current growth state, and $r_d$ the diet status. The output state summarizes the biological response of the herd, including body weight, average daily gain, feed intake, feed efficiency, heat production, and limiting factors. Thus, daily aggregation allows short-term climate decisions to be interpreted as biological consequences, while the growth update provides the information needed for feedback to the next climate-control decisions.

\subsection{Biological Guidance}

The outer loop returns the biological response of the herd to the climate optimizer as guidance for the next MPC decisions. This guidance is generated from the daily growth state, heat production, feed intake, and limiting factors:

\begin{equation}
b_d = G(g_d, \mathrm{HP}_d, \mathrm{FI}_d, L_d).
\label{eq:biological_guidance}
\end{equation}

This mapping transforms growth outputs into control guidance.

Here, $G(\cdot)$ is the guidance function and $g_d$ is the growth state. In implementation, this guidance can be expanded as a package:

\begin{equation}
\Gamma_d =
\left[
T^{\mathrm{low}}_{\mathrm{pref,d}},
T^{\mathrm{high}}_{\mathrm{pref,d}},
\lambda^{\mathrm{comfort}}_d,
\lambda^{\mathrm{energy}}_d,
\lambda^{\mathrm{vent/heat}}_d,
\beta^{\mathrm{bio}}_d
\right].
\label{eq:guidance_package_detail}
\end{equation}

The guidance package adjusts MPC preferences and decision priorities.

In this package, $T^{\mathrm{low}}_{\mathrm{pref,d}}$ lower comfort bound, $T^{\mathrm{high}}_{\mathrm{pref,d}}$ upper comfort bound, $\lambda^{\mathrm{comfort}}_d$ comfort weight, $\lambda^{\mathrm{energy}}_d$ energy weight, $\lambda^{\mathrm{vent/heat}}_d$ ventilation/heating priority, and $\beta^{\mathrm{bio}}_d$ biological bias. Thus, when body weight, heat production, feed response, or heat-stress risk changes, the same actuator candidates are evaluated with updated biological priorities. The guidance can modify the preferred comfort range, increase or decrease the comfort and energy weights, and shift ventilation or heating priorities. In SE-TDDT, MPC remains the final actuator-command selection mechanism, while SARG-SEL and the growth layer provide soft biological corrections rather than direct actuator commands.

\subsection{MPC Objective Function}

As in Eq.~\eqref{eq:mpc_objective}, at each five-minute control step, MPC evaluates the short-term climate rollout of each actuator candidate and assigns a decision cost to it. The cost function combines comfort violation, energy use, safety risk, actuator conflict, and biological benefit:

\begin{equation}
J(u_t^{(j)}) =
\alpha_c C_{\text{comfort}}^{(j)}
+\alpha_e C_{\text{energy}}^{(j)}
+\alpha_s C_{\text{safety}}^{(j)}
+\alpha_f C_{\text{conflict}}^{(j)}
-\beta_b B_{\text{bio}}^{(j)} .
\label{eq:mpc_objective_detail}
\end{equation}

The objective function evaluates each candidate actuator command.

Here, $\alpha_c,\alpha_e,\alpha_s,\alpha_f,\beta_b$ denote the decision weights. The biological term is subtracted because higher biological compatibility should reduce the final cost.

The final actuator command is selected by minimizing this cost over the candidate set in Eq.~\eqref{eq:control_candidates}:

\begin{equation}
u_t^* =
\arg\min_{u_t^{(j)} \in U_t}
J(u_t^{(j)}).
\label{eq:optimal_control_detail}
\end{equation}

The optimization selects the candidate command with the lowest cost.

In this relation, $j$ denotes the candidate index. Thus, MPC remains the final decision-maker: it does not directly follow biological guidance, but uses it as a soft correction while still respecting comfort, energy, safety, and operational-conflict constraints.

\subsection{Knowledge Components}

The knowledge layer of SE-TDDT converts single-episode training traces into compact knowledge usable for online edge control. Instead of storing all raw climate, actuator, growth, reward, and guidance records, the architecture organizes experience into four complementary components, as shown in Figure~\ref{fig:image_1}. CCLL-SEL is the local climate-context memory; it converts historical and local climate patterns into soft climate priors for MPC. SARG-SEL is the stage-aware biological guidance memory; it uses growth–diet reference paths to relate growth status, feed response, and biological stage to climate-control priorities. SS-KStore is the streaming memory; it stores step-wise training traces, actuator decisions, simulator responses, rewards, and daily summaries in a low-memory form. SETD-KStore is the final trans-domain knowledge package; it stores the policy snapshot, climate contexts, biological-guidance–decision relationships, reward indicators, coverage, and uncertainty for online control.

The four components can be summarized as

\begin{equation}
K_{\mathrm{TDDT}} =
\left\{
K^{\mathrm{climate}}_{\mathrm{CCLL}},
K^{\mathrm{bio}}_{\mathrm{SARG}},
K^{\mathrm{stream}}_{\mathrm{SS}},
K^{\mathrm{policy}}_{\mathrm{SETD}}
\right\}.
\label{eq:knowledge_components}
\end{equation}

This representation organizes operational knowledge for online control.

Here, $K_{\mathrm{TDDT}}$ is the knowledge set, $K^{\mathrm{climate}}_{\mathrm{CCLL}}$ the climate memory, $K^{\mathrm{bio}}_{\mathrm{SARG}}$ the biological guidance, $K^{\mathrm{stream}}_{\mathrm{SS}}$ the streaming memory, and $K^{\mathrm{policy}}_{\mathrm{SETD}}$ the policy package.

At the end of offline training, the final online-control package is constructed as

\begin{equation}
K_{\mathrm{SETD}}=\Psi(E_{\mathrm{train}},\mathrm{CCLL},\mathrm{SARG}).
\label{eq:knowledge_compression}
\end{equation}

This mapping compresses training experience into a loadable knowledge package.

Here, $K_{\mathrm{SETD}}$ denotes the final knowledge package, $\Psi$ the compression mapping, $E_{\mathrm{train}}$ the training experiences, CCLL the climate contexts, and SARG the biological references. During online deployment, the edge controller loads $K_{\mathrm{SETD}}$ instead of learning from scratch; MPC remains the final actuator-command selection mechanism, while the stored knowledge provides priors, guidance, and lightweight adaptation support.

\section{Dataset, Simulation Setup, and Edge Deployment}
\subsection{Barn and Herd Case Study}

The case study is a closed, 120-head cattle-fattening barn located at the Mahyar Agro-Industry and Dairy and Meat Complex in Qazvin Province, Iran. In SE-TDDT, this barn is treated as the Physical Twin, where sensors, actuators, outdoor weather, building specifications, energy leakage, and herd context provide the real input for simulation, calibration, command execution, and digital-twin updating. The geographical location is important because season, radiation, wind, outdoor temperature, and outdoor humidity directly affect barn thermal load, ventilation demand, and energy consumption.

The barn is represented as a physical-building entity, including walls, roof, openings, insulation, ventilation paths, and equipment placement. This information is transferred from the building information model (BIM) to the DT so that the climate simulator can account for the thermal and ventilation effects of the physical structure. The animals are modeled as a Holstein fattening herd, with breed assumptions related to growth, body weight, coat characteristics, carcass traits, and thermal response~\cite{kamal2014calving, calo1973growth, maia2005coat, ardicli2018retail}. The herd context includes the number of animals, average body weight, fattening day, and biological heat production. The IFC-based\footnote{Industry Foundation Classes (IFC) is an open standard for exchanging geometric and semantic building information.} BIM file used for the physical twin was designed in Blender\textsuperscript{\textregistered}.

The physical state of the barn is summarized as

\begin{equation}
P_{\mathrm{barn}}=\{\mathrm{BIM},H,S,A,W_{\mathrm{out}},L\}.
\label{eq:barn_state}
\end{equation}

This representation summarizes the physical inputs required for simulation.

\begin{figure}[htbp]
\centering
\includegraphics[width=0.95\linewidth,alt={Architecture connecting comfort, mechanistic climate, building information and growth models with the TDDT optimizer.}]{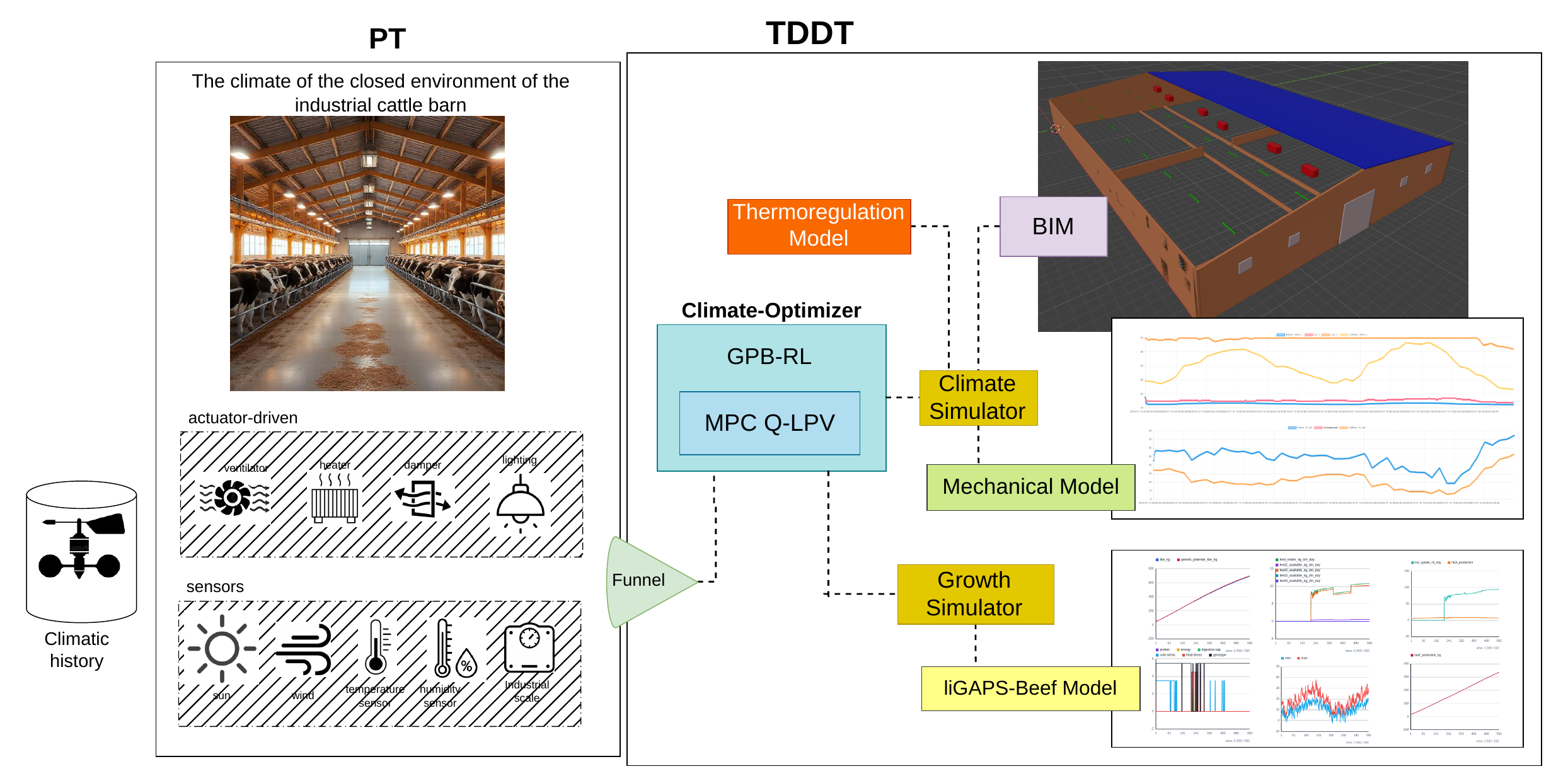}
\caption{Overview of the relationships among the comfort models, the mechanistic climate model, the building information model, and the growth model with the TDDT optimizer. Q-LPV denotes QLPV; the diagram label GPB-RL identifies the learning and tuning component.}
\label{fig:image_3}
\end{figure}

Here, $P_{\mathrm{barn}}$ denotes the barn state, BIM the building information model, $S$ the sensors, $A$ the actuators, $W_{\mathrm{out}}$ the outdoor weather, and $L$ the lighting and operation. The actuator set includes ventilation units, fans, dampers, heaters, lighting, and other controllable equipment used for heating, ventilation, and cooling management in controlled agricultural environments~\cite{worley2025greenhouses}. As shown in Figure~\ref{fig:image_3}, these physical, climatic, actuator, and herd components provide the input basis for the climate simulator, growth simulator, and TDDT optimizer. Sensor data are collected from the barn through the Funnel section. This is a GPIO\footnote{General-purpose input/output (GPIO) pins on a single-board computer are used for reading sensors or controlling actuators.} aggregator that receives data through the LoRa\footnote{LoRa (Long Range) is a low-power, long-range wireless communication technology.} server. 

\subsection{Climate Dataset and CCLL Construction}

The local climate dataset provides the contextual memory required by CCLL-SEL. In this study, 11 years of local outdoor climate data for the barn location were extracted from OpenWeather\textsuperscript{\textregistered}  for 2015–2025. The dataset includes outdoor temperature, relative humidity, wind speed and direction, radiation, pressure, rainfall, and, where available, indoor barn or nearby farm-station data. Its role is to prevent single-episode training from depending only on one simulated episode and to construct long-term regional climate priors for MPC.

The raw climate dataset, with wind direction (WD) as one of its variables, is represented as

\begin{equation}
Y=\{y_i\}_{i=1}^{N}, \qquad
y_i=(\tau_i,T_i,\mathrm{RH}_i,W_i,\mathrm{WD}_i,R_i,P_i,\mathrm{Rain}_i).
\label{eq:climate_observations}
\end{equation}

This representation defines the structure of the raw climate observations.

Here, $Y$ is the raw dataset, $y_i$ the climate record, $\tau_i$ the timestamp, $T_i$ the temperature, $\mathrm{RH}_i$ the relative humidity, $W_i$ the wind speed, $\mathrm{WD}_i$ the wind direction, $R_i$ the radiation, $P_i$ the pressure, and $\mathrm{Rain}_i$ the rainfall.

\begin{equation}
Y_{\mathrm{CCLL}}
=
Context\big(Aggregate(ResampleFill(Sync(Clean(Y))))\big).
\label{eq:climate_preparation}
\end{equation}

Here, $Y_{\mathrm{CCLL}}$ denotes the CCLL output; Clean, Sync, ResampleFill, Aggregate, and Context denote data cleaning, time alignment, resampling and gap filling, daily indicator aggregation, and context construction, respectively. Cleaning removes duplicate, invalid, unrealistic, or inconsistent records; synchronization places variables on a common temporal axis; resampling and gap filling convert the data to the required time step; aggregation generates daily and seasonal indicators. The final output is a set of climatic contexts and soft priors that allow CCLL-SEL to guide MPC using long-term local climate memory rather than only the current training episode.

\begin{figure}[htbp]
\centering
\includegraphics[width=0.95\linewidth,alt={Workflow for cleaning, synchronizing, aggregating and clustering historical climate data.}]{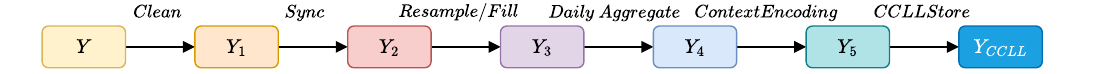}
\caption{CCLL-SEL data preparation workflow.}
\label{fig:ccll_preparation_workflow}
\end{figure}

Figure~\ref{fig:ccll_preparation_workflow} summarizes the construction of the CCLL-SEL dataset. In this workflow, \(Y\) is the raw 11-year local climate dataset, including timestamp, outdoor temperature, relative humidity, wind speed and direction, radiation, pressure, and rainfall. \(Y_1=Clean(Y)\) removes duplicate records, invalid times, physically impossible values, unrealistic jumps, and inconsistent observations. \(Y_2=Sync(Y_1,\Delta t)\) places all variables on a common temporal axis, and \(Y_3=ResampleFill(Y_2)\) converts the cleaned data to the required time step and fills short gaps by interpolation or local averaging. \(Y_4=Aggregate(Y_3)\) builds daily and seasonal climate indicators such as minimum, maximum, and mean temperature, mean humidity, mean wind speed, radiation sum, rainfall sum, and vapor-pressure deficit. \(Y_5=ContextEncode(Y_4)\) converts these indicators into climate-context descriptors such as season, heat level, humidity level, wind level, radiation level, and transition status. The final output \(Y_{\mathrm{CCLL}}=(c_d,k_d,q_d,p_d)_{d=1}^{D}\) is the CCLL contextual memory: \(c_d\) is the daily climate vector, \(k_d\) the nearest climatic context, \(q_d\) the context quality or confidence, and \(p_d\) the soft prior used by MPC to bias ventilation, heating, comfort, or energy decisions according to long-term local climate patterns.

\subsection{SARG Dataset and Guidance Construction}

\begin{figure}[t]
\centering
\includegraphics[width=0.95\linewidth,alt={Workflow mapping biological and diet-reference data to daily guidance and five-minute control features.}]{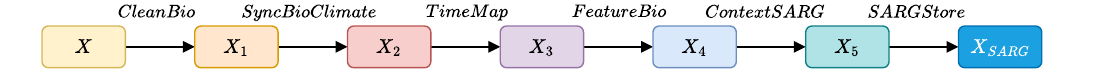}
\caption{SARG-SEL data preparation workflow. The raw biological and diet-reference dataset \(X\) is cleaned, synchronized with CCLL climate contexts, mapped from daily growth guidance to five-minute MPC steps, transformed into biological-control features, converted into SARG contexts, and stored as \(X_{\mathrm{SARG}}\) for biological guidance.}
\label{fig:sarg_workflow}
\end{figure}

Figure~\ref{fig:sarg_workflow} summarizes the construction of the SARG-SEL dataset. In this workflow, \(X\) is the raw biological and diet-reference dataset generated from the growth simulator, including body weight, ADG, feed intake, feed efficiency, heat production, limiting factors, biological stage, diet or reference program, and thermal-risk information. \(X_1=CleanBio(X)\) removes or flags biologically inconsistent values, such as negative weight, unrealistic ADG, impossible feed intake, abnormal feed efficiency, or heat production inconsistent with weight and feed. \(X_2=SyncBioClimate(X_1,Y_{\mathrm{CCLL}})\) synchronizes the cleaned biological trajectory with the daily climatic context from CCLL, so that the same growth or feed response can be interpreted differently under cold, warm, humid, hot, or transitional climate conditions. \(X_3=TimeMap(X_2,\Delta D=1\,day,\Delta t=5\,min)\) maps daily biological guidance to all five-minute MPC steps of the same day. \(X_4=FeatureBio(X_3)\) extracts biological-control features such as \(\mathrm{BW}_d\), \(\Delta \mathrm{BW}_d\), \(\mathrm{ADG}_d\), \(\mathrm{FI}_d\), \(\mathrm{FE}_d\), \(\mathrm{HP}_d\), limiting factor, diet reference, growth stage, and risk level, and compares them with reference paths. \(X_5=ContextSARG(X_4)\) converts each day into a biological-guidance context, including stage class, growth status, heat-load class, intake status, limitation class, diet class, climate match, best reference, top diet options, and confidence. The final output \(X_{\mathrm{SARG}}=(s_d,\phi^{\mathrm{bio}}_d,\epsilon^{ref}_d,\gamma_d,q^{\mathrm{bio}}_d)_{d=1}^{D}\) is the stage-aware SARG memory used to correct MPC through preferred comfort bounds, comfort/energy/gas weights, ventilation or heating priority, and biological bias.

\subsection{Growth-Simulator Inputs and Outputs}

The growth simulator represents the biological layer of SE-TDDT. Its role is to convert daily climate conditions, diet, herd status, and growth stage into biological indicators usable by the climate-control loop. In this work, the growth layer follows the LiGAPS-Beef logic, where climate, feed, breed, weight, and physiological state jointly affect animal growth, feed response, heat production, and biological limitations~\cite{vanderlinden2019ligaps1, vanderlinden2019ligaps2, vanderlinden2019ligaps3}.

The daily growth update is written as

\begin{equation}
y_d^{\mathrm{growth}}=F_g(\bar{x}_d,g_d,r_d,s_d).
\label{eq:growth_simulator}
\end{equation}

The growth model maps daily inputs to biological outputs.

Here, $F_g$ denotes the growth simulator, $g_d$ the herd status, $r_d$ the diet input, and $s_d$ the growth stage. The daily climate input may include minimum and maximum temperature, mean humidity, airflow, ventilation status, comfort violation, and energy use.

The output vector is

\begin{equation}
y_d^{\mathrm{growth}}=
[\mathrm{BW}_d,\mathrm{ADG}_d,\mathrm{FI}_d,\mathrm{FE}_d,\mathrm{HP}_d,L_d].
\label{eq:growth_simulator_output}
\end{equation}

The output vector summarizes the biological response of the herd.

These outputs explain how climate and diet affect animal performance and how herd growth changes future climate demand. For example, increased body weight and heat production raise barn thermal load, while reduced feed intake or feed efficiency may indicate climate, diet, digestion, energy, or protein limitation~\cite{vanderlinden2019ligaps1,vanderlinden2019ligaps2,vanderlinden2019ligaps3}. The growth outputs are therefore converted into biological guidance for MPC in the next climate-control cycle.

\subsection{Edge Platform}

To evaluate edge-level feasibility, the SE-TDDT framework was deployed on a low-power single-board computer. The platform was an Orange Pi 5\textsuperscript{\textregistered} equipped with an eight-core Arm processor, \qty{16}{\giga\byte} of random access memory (RAM), and a \qty{128}{\giga\byte} M.2 storage module. The operating system was Armbian 26.2.1 based on Debian Trixie. During training, the implementation used at most two central processing unit (CPU) cores and required up to \qty{4}{\giga\byte} of temporary memory.

The complete single-episode training was executed on this edge device. In the reported run, 5,761 training steps required 505,986 seconds. The 1000-day horizon comprises 288,000 nominal five-minute control intervals (Figure~\ref{fig:figure10}); the correspondence between the reported training-step count and these control intervals is not specified in the reported summary. After training, the edge controller loaded the final SETD-KStore knowledge package and executed online monitoring, decision support, and lightweight adaptation locally, without graphics processing unit (GPU) acceleration or cloud-side computation. These reported results establish computational feasibility; field validation with real barn data remains future work. Although the offline training time was long, executing the full workflow on an edge device is valuable because it demonstrates that trans-domain knowledge can be generated, stored, and later reused for barn-level control under low-power farm-computing constraints.

\subsection{Single-Episode Training Workflow}

The SE-TDDT training workflow links the five-minute inner climate loop with the daily outer growth loop over a 1000-day fattening horizon. This use of a single long training episode is inspired by episodic control, where stored past experience can support future decisions, but in SE-TDDT the final actuator decision remains constrained by MPC~\cite{blundell2016episodic}. During this single episode, the climate simulator generates short-term barn responses for actuator decisions, while MPC evaluates candidate commands under comfort, energy, safety, and biological-priority constraints. The resulting five-minute climate records are then aggregated into daily climate summaries and transferred to the growth simulator.

The daily growth simulator computes biological outputs, including body weight, average daily gain, feed intake, feed efficiency, heat production, and limiting factors. These outputs are compared with SARG reference patterns to generate biological guidance for the next climate-control decisions. Thus, the training process does not only simulate climate or growth separately; it converts their interaction into trans-domain experience.

The knowledge generated by the two loops is stored through a hierarchical memory path:

\[
e^{\mathrm{in}}_t, e^{\mathrm{out}}_d \rightarrow \mathrm{SS}\text{-}KStore \rightarrow \mathrm{SETD}\text{-}KStore.
\]

Here, $e^{\mathrm{in}}_t$ denotes the five-minute climate-control record, including state, actuator decision, cost, guidance, and decision weights, while $e^{\mathrm{out}}_d$ denotes the daily biological record, including aggregated climate, growth output, biological guidance, biological context, and data quality. SS-KStore stores these records in a streaming form during training, and SETD-KStore compresses them into a final loadable knowledge package for online edge control.

\begin{figure}[htbp]
    \centering
    \includegraphics[width=0.99\linewidth,alt={Single-episode training workflow linking five-minute decisions, daily growth guidance, streaming memory and compressed knowledge.}]{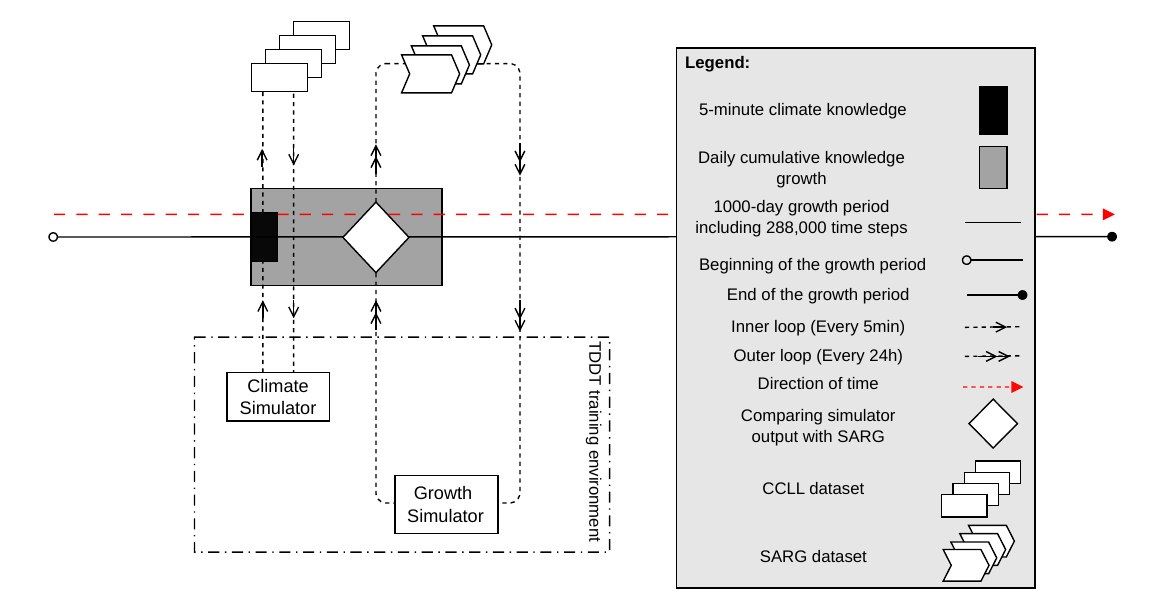} 
    \caption{Schematic of SE-TDDT single-episode training over a 1000-day fattening horizon. The inner loop generates 5-minute climate-control knowledge, which is aggregated into daily climate summaries for the outer growth loop. The growth simulator computes daily biological outputs, compares them with SARG reference patterns, and returns biological guidance to the optimizer. The resulting training experience is stored first in SS-KStore and then compressed into SETD-KStore for online edge control.}
    \label{fig:figure10}
\end{figure}

Figure~\ref{fig:figure10} summarizes this single-episode workflow. It shows how 5-minute climate knowledge is generated in the inner loop, transformed into daily cumulative knowledge for the growth loop, synchronized over the 1000-day fattening period, compared with CCLL and SARG references, and finally converted into reusable knowledge for SS-KStore and SETD-KStore.

\section{Verification, Validation, and Experimental Design}

\subsection{Verification of Model Coupling}

The first verification step examines whether the technical coupling among the SE-TDDT components is continuous, ordered, and traceable after the 1000-day single-episode training. The verified computational path is

\[
\begin{gathered}
\text{Climate Simulator} \rightarrow \text{MPC} \rightarrow \text{Daily Aggregation} \\
\rightarrow \text{Growth Simulator} \rightarrow \text{SARG-SEL} \rightarrow \text{MPC}.
\end{gathered}
\]

This pathway represents the climate--growth decision coupling examined during verification.

The memory path is verified as

\[
\text{SS-KStore}
\rightarrow
\text{SETD-KStore}.
\]

This pathway represents the transfer of knowledge from training to control.

The verification uses five-minute climate traces, daily growth outputs, MPC decision records, biological-guidance records, SARG-SEL contexts, CCLL-SEL contexts, and KStore artifacts. These reports allow traceability across climate, growth, decision, guidance, and memory. In particular, the verification checks whether temporal rates are compatible, data ordering is preserved, climate and growth units remain consistent, guidance is returned to MPC, and training records are reconstructed correctly in the KStore components. The CCLL-SEL connection is also verified through the reconstruction of 4018 climatic descriptors and 12 contextual centroids; mapping the growth trajectory to climate contexts confirms that the outer loop is connected not only to raw daily climate data but also to climatic contextual memory, as shown in Figure~\ref{fig:ccll_clusters}. Therefore, the verification confirms that SE-TDDT has a technically consistent coupling structure and can proceed to simulator validation, closed-loop validation, baseline comparison, and ablation study.

\subsection{Closed-Loop Validation}

Closed-loop validation evaluates whether SE-TDDT works as an executable bio-aware control cycle, rather than as a loose connection between separate simulators. In the validation design, the complete evaluation path is defined as

\[
\begin{gathered}
\text{Verification} \rightarrow \text{Simulator Validation} \rightarrow \text{Closed-Loop Validation} \\
\rightarrow \text{Baseline Comparison} \rightarrow \text{Ablation Study}.
\end{gathered}
\]

This sequence defines the overall evaluation procedure.

Verification checks the technical connection among the climate simulator, growth simulator, MPC, CCLL-SEL, SARG-SEL, SS-KStore, and SETD-KStore. Simulator validation examines the independent behavior of the climate and growth simulators. Closed-loop validation then evaluates the integrated climate–growth–MPC–guidance cycle. Baseline comparison compares SE-TDDT with simpler configurations, and ablation study isolates the contribution of each knowledge or adaptation component.

The compact closed-loop path is

\[
x_t \rightarrow \bar{x}_d \rightarrow y_d^{\mathrm{growth}} \rightarrow b_d \rightarrow \mathrm{MPC} .
\]

This pathway describes the biological feedback provided to MPC.

\begin{figure}[htbp]
\centering
\includegraphics[width=0.95\linewidth,alt={Sankey diagram linking climate context, biological guidance, controller decisions and outcomes.}]{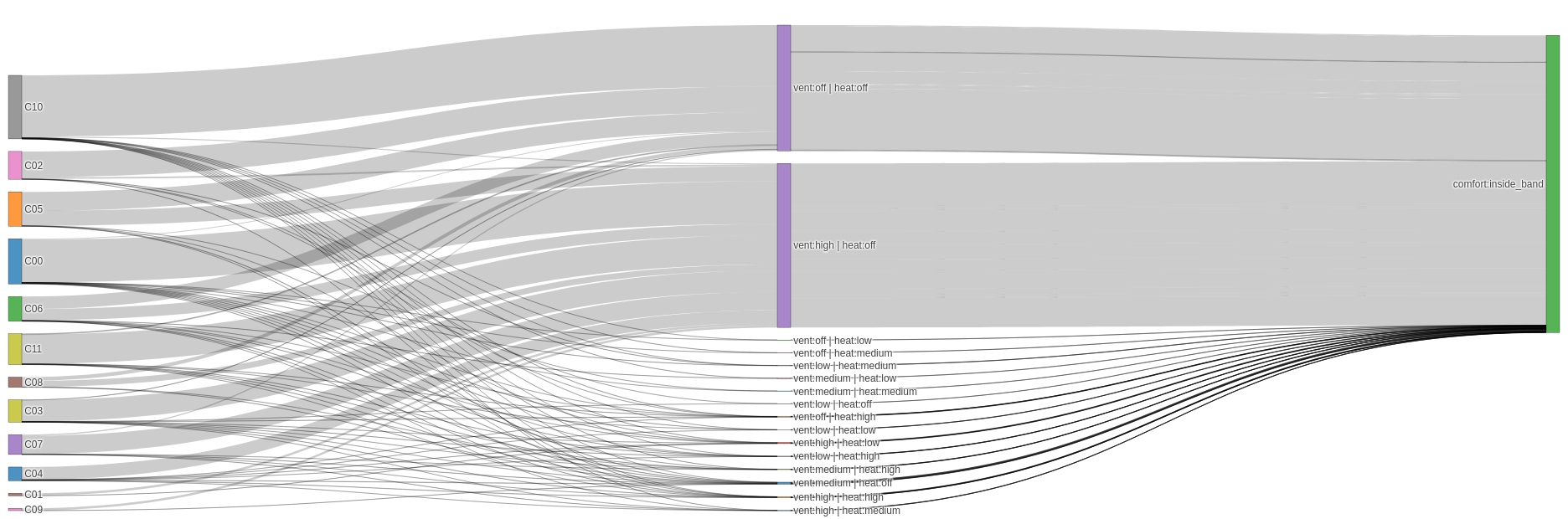}
\caption{Trans-domain decision-flow structure. The Sankey diagram summarizes how climate context, biological guidance, MPC decisions, and outcomes are interconnected within the TDDT pipeline. It also illustrates the relationship between CCLL climate contexts and SARG biological guidance in supporting MPC decision-making.}
\label{fig:image_12_7}
\end{figure}

In this path, five-minute climate data are aggregated into daily indicators, the growth simulator generates biological outputs, SARG-SEL converts them into guidance, and MPC uses that guidance in the next climate decision. Figure~\ref{fig:image_12_7} shows this climate–growth–guidance–MPC cycle.

The reported single-episode execution exhibited consistent closed-loop behavior over the evaluated simulation period: the climate loop kept the barn within the comfort range with a comfort-violation rate of 0\%, while the growth loop produced a 1000-day biological trajectory with a final weight of about \qty{917}{\kilogram}, growth accuracy of 96.91\%, and a coefficient of determination of $R^2 \approx 0.996$. However, the validation also shows that performance was not ideal in all dimensions: energy accuracy was about 92.61\%, while low feed-consumption accuracy remained weak at about 51\%. Therefore, the closed loop confirms executable bio-aware coupling, but feed-pressure reduction still requires stronger SARG-SEL tuning and better feed–growth trade-off weighting.

\subsection{Baseline and Ablation Design}

The baseline design is used to separate the contribution of prediction, growth feedback, and knowledge memory in the proposed control framework. The comparison includes four scenarios:

\begin{equation}
B=\{B_0,B_1,B_2,B_3\}.
\label{eq:baselines}
\end{equation}

This set defines the baseline evaluation scenarios.

Here, $B$ denotes the baseline set, $B_0$ simple HVAC control, $B_1$ climate MPC, $B_2$ TDDT without memory, and $B_3$ the full SE-TDDT configuration. In $B_0$, simple HVAC control uses only temperature or ventilation thresholds and has no future prediction, growth feedback, or knowledge prior. In $B_1$, MPC predicts the barn climate response and supports climate-aware decisions, but biological outputs such as heat production, average daily gain, feed intake, feed efficiency, and limiting factors do not return to the decision function. In $B_2$, climate–growth coupling exists, but knowledge-memory support is removed or limited; this scenario evaluates how the absence of memory affects adaptation and reuse of single-episode experience. In $B_3$, the complete SE-TDDT configuration is used as the reported full mode.

The current full SE-TDDT results reported for $B_3$ are:

\begin{equation}
\mathrm{Comfort}=100,\quad \mathrm{Energy}=92.61,\quad \mathrm{Growth}=96.91,\quad \mathrm{Feed}=51.00.
\label{eq:reported_performance}
\end{equation}

These indicators summarize the reported reference performance of the full configuration.

These scenarios provide a compact ablation design because each baseline removes one level of capability: reactive HVAC control, predictive climate control without growth feedback, climate–growth coupling without memory, and the full trans-domain memory-supported framework. Therefore, the comparison can show whether the observed performance is mainly caused by simple regulation, MPC prediction, growth feedback, or the complete SE-TDDT architecture.

\section{Results}

\subsection{Climate Comfort and Safety Results}

The inner-loop validation shows that SE-TDDT preserved the barn climate within the animal comfort range during the reported tracking period. The thermal-comfort indicator reached 100.00\%, and the comfort-violation rate was reported as

\begin{equation}
R_{\mathrm{viol}}=0.0000 .
\label{eq:violation_rate}
\end{equation}

The reported comfort-violation rate is zero.

Here, $R_{\mathrm{viol}}$ denotes the comfort-violation rate. Figure~\ref{fig:integrated_results_a} shows that the simulated indoor temperature remained within the comfort limits without any comfort violation.

Safety was evaluated through actuator behavior, especially the heating–ventilation conflict. After applying the safety filter, the heating–ventilation conflict rate was

\begin{equation}
C_{\mathrm{heat-vent}}=0.00 ,
\label{eq:conflict_rate}
\end{equation}

which indicates no heating–ventilation conflict.

Here, $C_{\mathrm{heat-vent}}$ denotes the heating--ventilation conflict rate. The reported actuator-safety indicator was as follows:

\begin{equation}
\mathrm{Safety}_{\mathrm{actuator}}=96.96 .
\label{eq:actuator_safety}
\end{equation}

This indicator summarizes the reported actuator-safety performance.

Here, $\mathrm{Safety}_{\mathrm{actuator}}$ denotes the reported actuator-safety indicator. These values show that the interlock mechanism limited heating during intense ventilation and kept the actuator decisions executable. Table~\ref{tab:Practical_control_penalties} supports this result by reporting practical control penalties during single-episode learning, including reward mean (0.8403), comfort violation rate (0.0000), and conflict penalty mean (0.0028).

\begin{table}[htbp]
\centering
\caption{Practical control penalties during single-episode learning.}
\label{tab:Practical_control_penalties}
\begin{tabular}{@{}lr@{}}
\toprule
Metric & Value \\
\midrule
Reward mean & 0.8403 \\
Comfort violation rate & 0.0000 \\
Conflict penalty mean & 0.0028 \\
\bottomrule
\end{tabular}
\end{table}

However, the actuator results also show that safety did not fully imply smooth operation. The report includes 2484 command changes, 695 sudden changes, and 246 reversal oscillations. Therefore, the climate loop confirmed comfort preservation and primary safety, but actuator smoothing remains an implementation limitation. Figures~\ref{fig:integrated_results_b} and~\ref{fig:integrated_results_c} summarize energy and actuator utilization; the reported switching and oscillation counts are given in Eq.~\eqref{eq:actuator_events}.

\subsection{Growth and Biological Response}

The outer-loop results show that the growth simulator converted daily aggregated climate, herd status, and diet into biological responses at the daily scale. The biological output vector is

\begin{equation}
y_d^{\mathrm{growth}}=(\mathrm{BW}_d,\mathrm{ADG}_d,\mathrm{FI}_d,\mathrm{FE}_d,\mathrm{HP}_d,L_d).
\label{eq:daily_growth_output}
\end{equation}

This vector summarizes the daily biological response of the herd.

These variables are the main biological indicators returned from the growth layer to the SE-TDDT decision process.

The reported closed-loop execution produced a 1000-day growth trajectory, with final body weight reaching about \qty{917}{\kilogram} and growth accuracy reported as 96.91\%, with $R^2 \approx 0.996$. This indicates that the simulated body-weight trajectory followed the reference growth path closely. ADG was used to evaluate daily growth dynamics, while feed intake and feed efficiency were used to detect whether climate and diet decisions supported efficient biological performance. Heat production acted as the main biological feedback to the climate loop because increasing body weight and metabolic heat modify the barn thermal load and future ventilation demand~\cite{vanderlinden2019ligaps1, vanderlinden2019ligaps2, vanderlinden2019ligaps3}.

The dominant limiting factors were interpreted through the growth model outputs, especially feed-related and climate-related constraints. The results indicate that growth tracking was strong, but feed performance remained weaker than the other dimensions; low feed-consumption accuracy was reported at about 51.00\%. Therefore, the biological response confirms that SE-TDDT can connect climate decisions to growth, feed, heat production, and limiting factors, but also shows that feed-pressure management requires further improvement.

\subsection{Energy and Actuator Behavior}

Energy and actuator behavior were evaluated to determine whether comfort preservation was achieved with executable and stable control commands. At the energy level, the main criterion was controllable actuator consumption, reported through the low-energy-consumption indicator. The full SE-TDDT execution achieved

\begin{equation}
\mathrm{Acc}_{\mathrm{energy}}=92.61\%.
\label{eq:energy_performance}
\end{equation}

This indicator reports energy-related performance.

Here, $\mathrm{Acc}_{\mathrm{energy}}$ denotes the reported energy-accuracy metric. Figure~\ref{fig:integrated_results_b} shows the interaction among climate pressure, controllable energy consumption, and comfort error during inner-loop climate control.

Actuator behavior was evaluated through ventilation/heating intensity and command stability. Figure~\ref{fig:integrated_results_c} shows that the inner loop was dominated by ventilation, while heating remained low; this indicates that the controller preserved comfort mainly through ventilation control rather than intensive heating. Safety was also supported by the absence of heating–ventilation conflict after filtering, but command smoothness remained incomplete.

The actuator switching summary is

\begin{equation}
N_{\mathrm{change}}=2484,\quad
N_{\mathrm{sudden}}=695,\quad
N_{\mathrm{reverse}}=246.
\label{eq:actuator_events}
\end{equation}

These event counts characterize the limitations in actuator smoothness.

Here, $N_{\mathrm{change}}$ denotes the number of command changes, $N_{\mathrm{sudden}}$ the number of sudden changes, and $N_{\mathrm{reverse}}$ the number of reversal events. These values show that, although SE-TDDT preserved comfort and controlled energy use, the actuator policy still produced frequent changes, sudden transitions, and reversal oscillations. Therefore, the energy result is acceptable, but future optimization should add stronger switching penalties, ramp-rate limits, or smoother actuator constraints to reduce oscillation and improve practical deployability. The event counts in Eq.~\eqref{eq:actuator_events} support this limitation.

\subsection{Knowledge Transfer and Online Readiness}

The knowledge-transfer result shows how SE-TDDT converts offline single-episode training into a loadable online-control package. During training, SS-KStore acts as the streaming memory layer. It stores climate states, actuator decisions, simulator responses, rewards, daily biological guidance, and temporal summaries in an append-only and low-memory form. This prevents the edge controller from keeping the complete high-frequency training trajectory in memory.

At the end of training, SETD-KStore converts the stored experience into the final trans-domain knowledge package:

\begin{equation}
K_{\mathrm{SETD}}=\Psi(E_{\mathrm{train}},\mathrm{CCLL},\mathrm{SARG}).
\label{eq:knowledge_compression_result}
\end{equation}

This mapping converts training experience into reusable online-control knowledge.

Here, CCLL denotes the climate contexts and SARG the biological references. The resulting package contains the policy snapshot, climate contexts, biological-guidance–decision relationships, reward indicators, coverage, and uncertainty.

The online deployment path is summarized as

\[
K_{\mathrm{SETD}} \rightarrow \text{Edge Controller} \rightarrow \text{MPC}.
\]

This pathway describes the transfer of learned knowledge to online control.

Here, the edge controller is the local runtime component, and MPC selects the actuator commands. This means that online control does not start from scratch. Instead, the edge controller loads the offline SE-TDDT knowledge, uses it as prior guidance for MPC, and is designed to perform lightweight adaptation with real barn data during future field deployment. Therefore, SS-KStore connects training traces to knowledge extraction, while SETD-KStore connects offline single-episode learning to online edge deployment. As shown conceptually in Figure~\ref{fig:image_1}, this memory path is the mechanism that makes SE-TDDT usable beyond simulation as a deployable barn-level control architecture.

\subsection{Main Quantitative Summary}

The root mean square error (RMSE) is reported relative to the mean in the growth-fit metric. The main quantitative results indicate that SE-TDDT is executable as a trans-domain climate–growth–energy–feed control cycle. The strongest result is thermal-comfort preservation, followed by a coherent long-term growth trajectory and acceptable energy behavior. The main performance limitations are feed performance and actuator smoothness (Table~\ref{tab:metrics_performance}).

\begin{table}[h]
\centering
\caption{System Performance Metrics and Interpretation}
\label{tab:metrics_performance}
\small
\begin{tabularx}{\linewidth}{@{}P{2.8cm} P{4.0cm} Y@{}}
\toprule
\textbf{Metric} & \textbf{Value} & \textbf{Interpretation} \\ Thermal comfort & 100.00\% & The controller preserved the comfort constraint. \\
\midrule
Comfort violation & 0.0000 & No comfort violation was reported. \\
Growth/production & 96.91\% & The growth trajectory remained close to the reference path. \\
Growth fit & RMSE/mean = 0.031; $R^2 = 0.995745$ & The biological trajectory was continuous and coherent. \\
Energy performance & 92.61\% & Energy behavior was acceptable, but requires further improvement. \\
Feed performance & 51.00\% & Feed pressure remained the main weakness. \\
Actuator smoothness & 2484 changes; 695 sudden changes; 246 reversals & Command smoothing requires further improvement. \\ \bottomrule
\end{tabularx}
\end{table}

\begin{figure}[p]
\centering

\begin{subfigure}[t]{0.45\textwidth}
    \centering
    \includegraphics[width=\linewidth,height=0.22\textheight,keepaspectratio,alt={Time series of indoor and outdoor temperature and lower and upper comfort bounds.}]{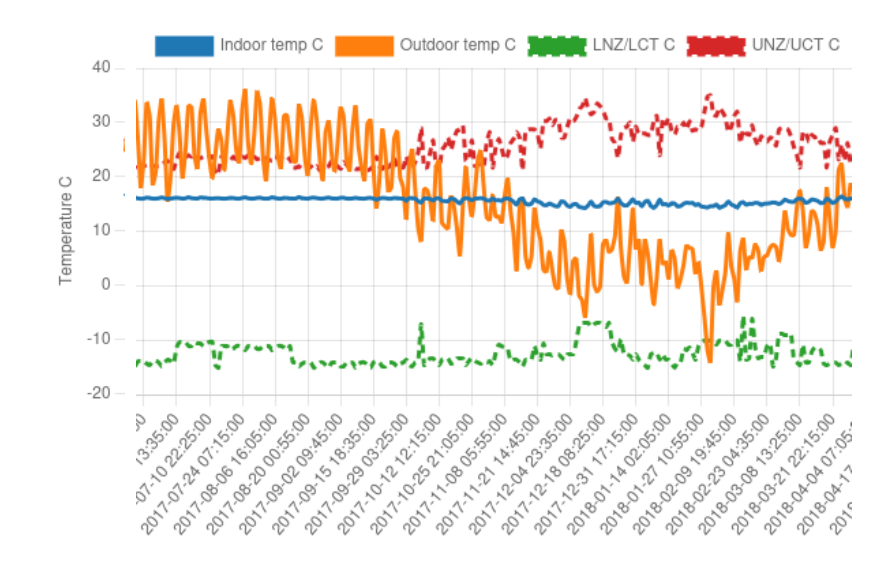}
    \caption{Climate comfort tracking.}
    \label{fig:integrated_results_a}
\end{subfigure}
\hfill
\begin{subfigure}[t]{0.45\textwidth}
    \centering
    \includegraphics[width=\linewidth,height=0.22\textheight,keepaspectratio,alt={Three-dimensional surface of comfort error versus outdoor temperature and actuator energy.}]{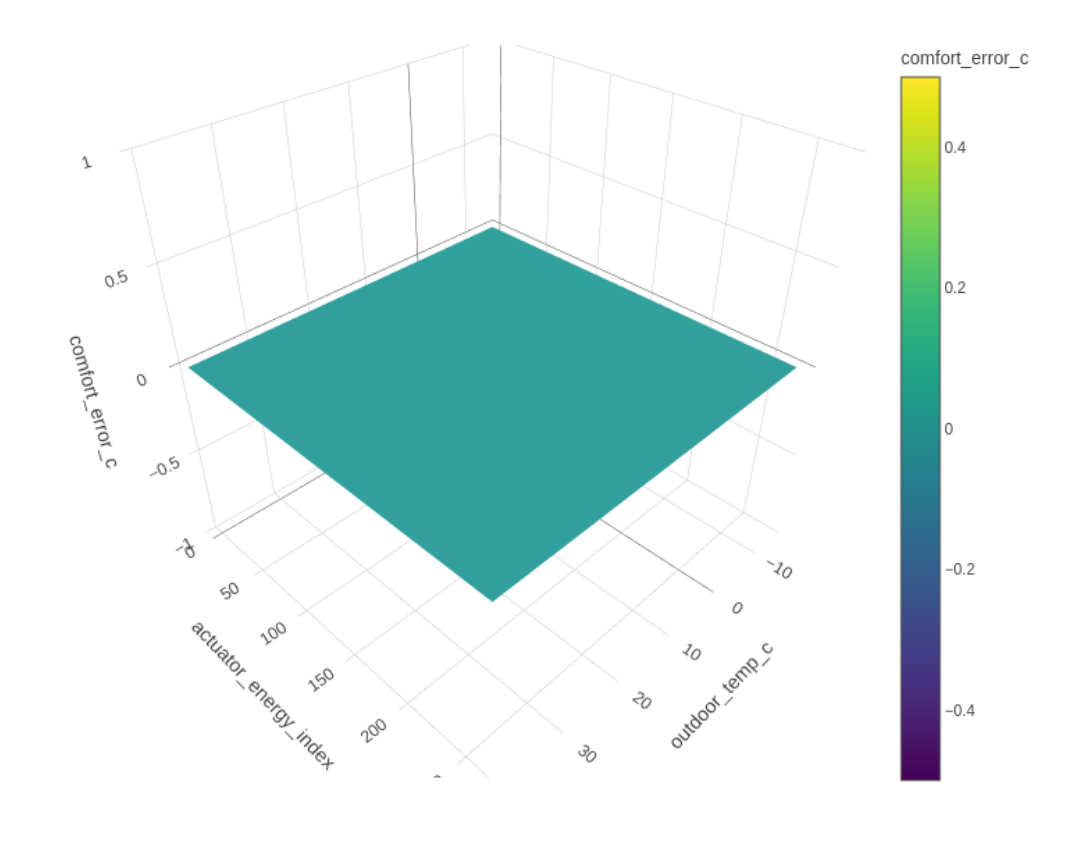}
    \caption{Climate--energy--comfort response.}
    \label{fig:integrated_results_b}
\end{subfigure}

\vspace{0.6em}

\begin{subfigure}[t]{0.45\textwidth}
    \centering
    \includegraphics[width=\linewidth,height=0.22\textheight,keepaspectratio,alt={Time series of ventilation and heating intensity.}]{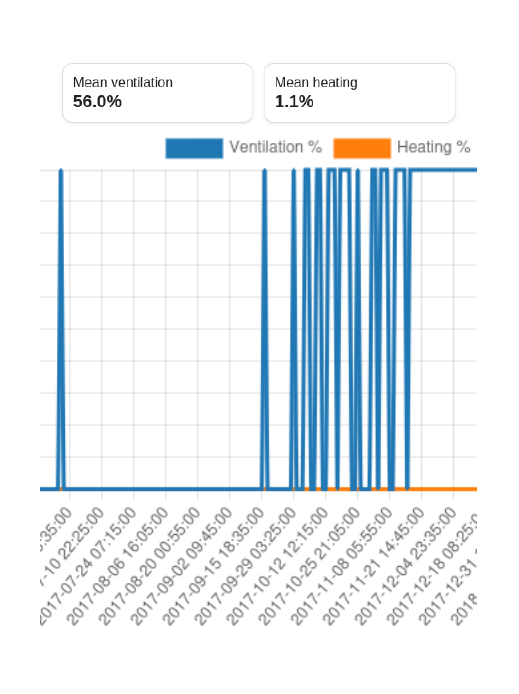}
    \caption{Ventilation and heating intensity.}
    \label{fig:integrated_results_c}
\end{subfigure}
\hfill
\begin{subfigure}[t]{0.45\textwidth}
    \centering
    \includegraphics[width=\linewidth,height=0.22\textheight,keepaspectratio,alt={Three-dimensional MPC score surface across heating and ventilation operating points.}]{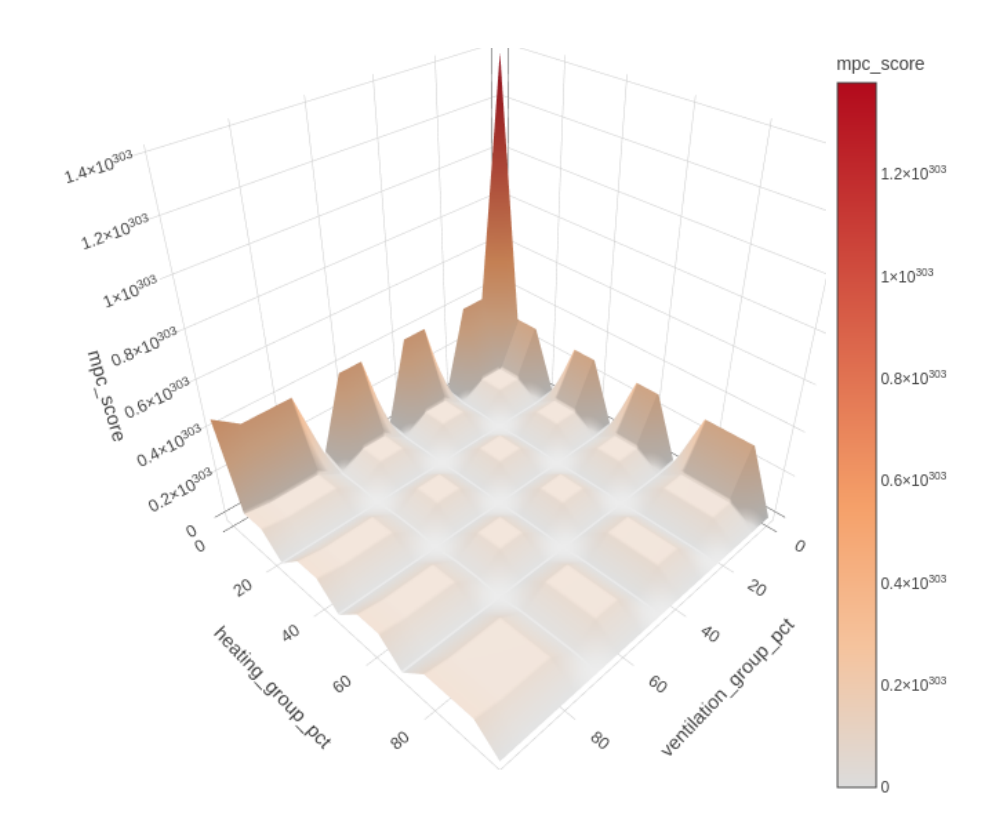}
    \caption{MPC score versus heating and ventilation intensity.}
    \label{fig:integrated_results_d}
\end{subfigure}

\vspace{0.6em}

\begin{subfigure}[t]{0.45\textwidth}
    \centering
    \includegraphics[width=\linewidth,height=0.22\textheight,keepaspectratio,alt={Radar plot comparing biological and feed indicators across five fattening phases.}]{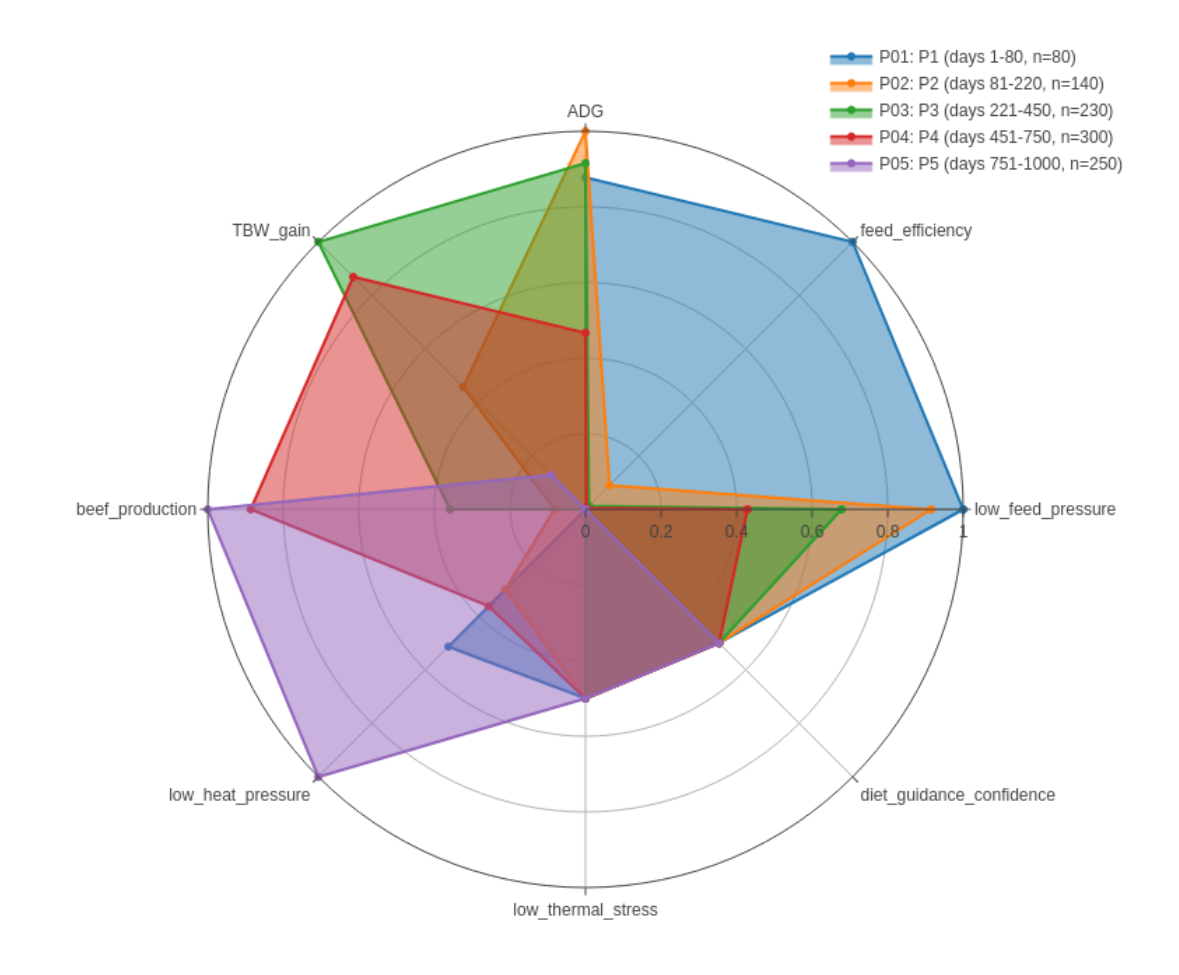}
    \caption{Phase-based feed--growth radar comparison.}
    \label{fig:integrated_results_e}
\end{subfigure}
\hfill
\begin{subfigure}[t]{0.45\textwidth}
    \centering
    \includegraphics[width=\linewidth,height=0.22\textheight,keepaspectratio,alt={Three-dimensional surface of body-weight response across growth phases and feed intake.}]{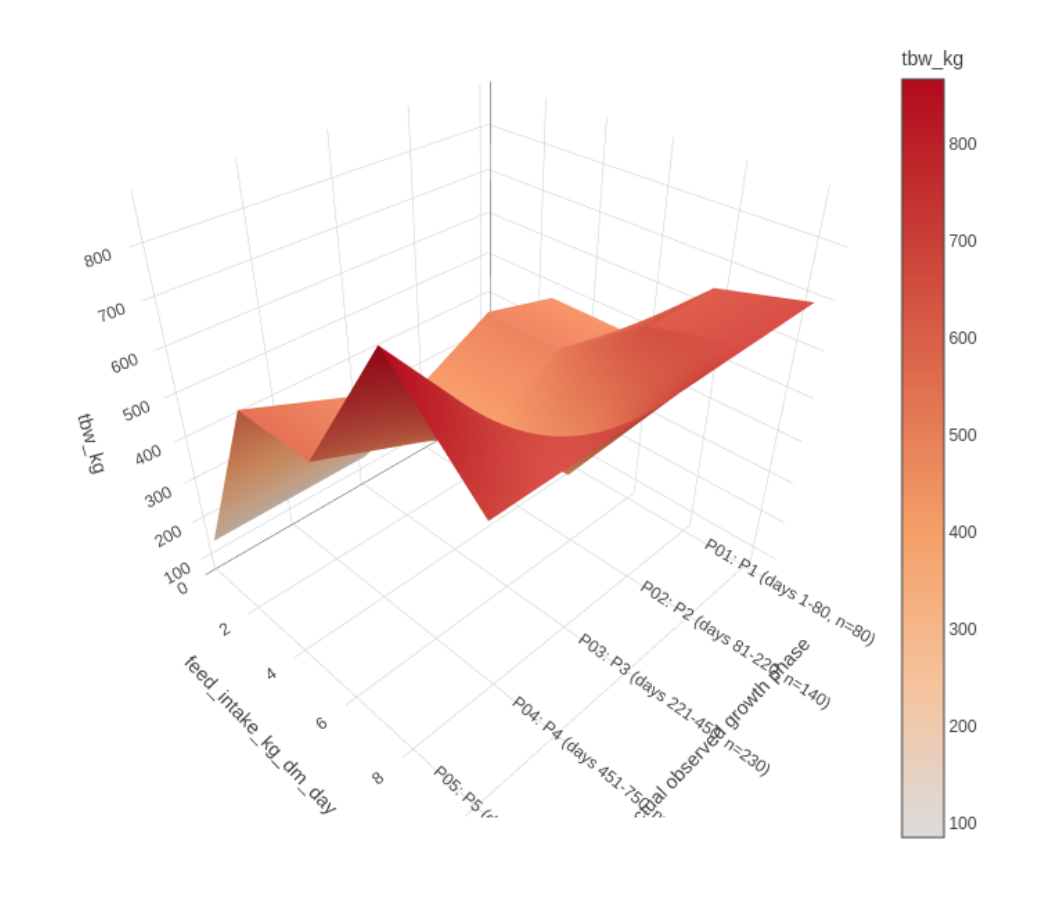}
    \caption{Feed--growth--production response.}
    \label{fig:integrated_results_f}
\end{subfigure}

\caption{
Integrated SE-TDDT performance summary.
(a) Inner-loop climate tracking showing indoor temperature within the animal comfort bounds.
(b) Climate--energy--comfort response showing the interaction among climate pressure, controllable energy consumption, and comfort error.
(c) Ventilation and heating intensity, indicating how comfort was maintained by actuator operation.
(d) MPC score surface across heating and ventilation operating points.
(e) Radar comparison of biological and feed indicators across the five fattening phases.
(f) Body-weight response across growth phases and feed intake, illustrating the feed--growth--production relationship.
}
\label{fig:integrated_results}
\end{figure}

Overall, the results show that adding biological feedback, knowledge memory, and adaptive tuning preserved the primary climate-safety requirement: thermal comfort was preserved while growth feedback was returned to the climate-control loop. However, the reported low feed-consumption accuracy of 51.00\% shows that the achieved growth was accompanied by higher feed pressure than the SARG reference. In addition, actuator behavior still included frequent switching, sudden changes, and reversal events. Therefore, the reported results demonstrate the simulation-based feasibility of executable trans-domain coupling, while future work should improve feed-pressure management and actuator smoothness. Figure~\ref{fig:integrated_results_c} reports ventilation/heating behavior, Figure~\ref{fig:integrated_results_b} summarizes the climate–energy–comfort response, Figure~\ref{fig:integrated_results_d} shows the MPC score surface, Eq.~\eqref{eq:actuator_events} reports actuator switching and oscillation, and Figure~\ref{fig:integrated_results_e} and Figure~\ref{fig:integrated_results_f} support the interpretation of growth, feed, heat production, and feed-pressure behavior.

\subsection{Comparison with Holstein Fattening Literature}

The growth and feed results were also compared with reported Holstein and Holstein-Friesian fattening studies. This comparison should be interpreted carefully because the SE-TDDT simulation follows an extended 1000-day fattening horizon, whereas many reference studies report conventional slaughter endpoints or shorter feedlot windows. Therefore, direct day-by-day comparison is not always appropriate; the most meaningful comparison is based on matched growth windows or similar live-weight endpoints~\citep{walter2016,kim2021,kim2023}.

SE-TDDT reached a final body weight of \qty{917.1}{\kilogram}, which is higher than reported Holstein slaughter body-weight values of about \qtyrange{745.6}{820.3}{\kilogram} in Kim et al.~\citep{kim2023} and up to \qty{812.8}{\kilogram} in Kim et al.~\citep{kim2021}. Mean ADG was \qty{0.875}{\kilogram\per\day}, lower than the \qtyrange{1.12}{1.26}{\kilogram\per\day} range reported by Kim et al.~\citep{kim2023} and lower than several 28-day Holstein feedlot windows reported by Walter et al.~\citep{walter2016}. Mean dry matter intake (DMI) was \qty{7.93}{\kilogram\per\day}, which is close to common Holstein feedlot values of about \qtyrange{8.2}{11.98}{\kilogram\per\day} reported in Holstein feeding references~\citep{walter2016,kim2023,iowa2020}. Feed efficiency decreased in the late stage, which is consistent with reports showing that longer feeding increases body weight but reduces gain efficiency~\citep{walter2016,kim2021}. Therefore, the final SE-TDDT stage should be interpreted as an extended fattening scenario rather than a conventional slaughter endpoint (Table~\ref{tab:holstein_result_comparison}).

\begin{table}[t]
\centering
\small
\caption{Compact comparison of SE-TDDT growth and feed outputs with Holstein fattening literature.}
\label{tab:holstein_result_comparison}
\begin{tabularx}{\linewidth}{@{}P{2.5cm} P{2.2cm} Y Y@{}}
\toprule
Metric & SE-TDDT output & Literature range / reference & Interpretation \\
\midrule
Final BW 
& \qty{917.1}{\kilogram} 
& About \qtyrange{745.6}{820.3}{\kilogram}~\citep{kim2023}; up to \qty{812.8}{\kilogram}~\citep{kim2021} 
& Higher final BW due to the extended 1000-day horizon. \\

Mean ADG 
& \qty{0.875}{\kilogram\per\day} 
& About \qtyrange{1.12}{1.26}{\kilogram\per\day}~\citep{kim2023}; \qtyrange{1.02}{1.77}{\kilogram\per\day} in 28-day windows~\citep{walter2016} 
& Growth is coherent but conservative in several windows. \\

Mean DMI 
& \qty{7.93}{\kilogram\per\day} 
& About \qtyrange{8.2}{11.98}{\kilogram\per\day}~\citep{walter2016,kim2023,iowa2020} 
& Feed intake is close to reported feedlot values. \\

Feed efficiency 
& Declines in late stage 
& Longer feeding reduces gain efficiency~\citep{walter2016,kim2021} 
& Late-stage decline is biologically plausible. \\

Beef-production proxy 
& \qty{487.4}{\kilogram} 
& Carcass weights about \qtyrange{417.1}{458.7}{\kilogram}~\citep{kim2023} 
& Proxy is higher because the simulation extends beyond conventional endpoints. \\
\bottomrule
\end{tabularx}
\end{table}

Overall, the literature comparison supports the plausibility of the simulated growth and feed trajectories, but it also confirms that the final phase of SE-TDDT should be interpreted as an extended fattening horizon. Agreement with the published literature is strongest for BW, ADG, DMI, and feed-efficiency trends, whereas heat production, comfort violation, actuator commands, and energy consumption remain mainly internal mechanistic or control-oriented outputs.

\FloatBarrier
\section{Discussion}

\subsection{Interpretation of Main Findings}

The main finding is that SE-TDDT demonstrates the feasibility of transforming livestock-building climate control from a single-domain HVAC problem into a bio-aware and trans-domain decision problem. The reported results show that climatic comfort was preserved while a coherent 1000-day growth trajectory was generated, confirming that climate, growth, feed, energy, biological guidance, memory, and MPC can remain connected within one executable control cycle. This means that adding biological feedback and knowledge memory preserved the primary climate-safety requirement; instead, it allowed the controller to include livestock response while preserving comfort.

The growth loop is not only a reporting layer. It converts daily aggregated climate data into biological outputs such as growth, feed response, heat production, and limiting factors, and returns them to MPC as guidance. Therefore, MPC remains the final actuator-command selection mechanism, but its candidate evaluations are corrected by biological priorities such as comfort-range adjustment, ventilation/heating priority, energy weight, and sensitivity to heat production or limiting factors.

The memory layer explains how offline experience becomes usable in online control. SS-KStore stores training traces in a streaming form, while SETD-KStore compresses the single-episode experience into a loadable package containing policy snapshots, climate contexts, biological-guidance–decision relationships, reward indicators, coverage, and uncertainty. At runtime, the edge controller loads this package instead of learning from scratch and is designed to fine-tune decisions with real barn data during future field deployment. Thus, the main interpretation is that SE-TDDT is not only a simulation pipeline, but a deployable architecture for transferring offline trans-domain knowledge to online barn-level control.

\subsection{Coverage-Based Comparison with Existing Approaches}

To clarify the position of SE-TDDT relative to existing approaches, Table~\ref{tab:coverage_comparison} compares the main dimensions covered by representative study groups. The comparison is not intended to rank all methods numerically; instead, it shows which domains are connected within one executable feedback loop. Existing agricultural DTs commonly support monitoring, prediction, and environmental management, while MPC and Bayesian optimization studies mainly focus on climate comfort, uncertainty, and energy. Livestock-building climate studies are closer to animal housing, but usually do not return long-term growth, feed efficiency, heat production, or biological limitation to the actuator decision. Mechanistic growth models such as LiGAPS-Beef represent growth, feed, heat production, and limiting factors, but they are not climate-control optimizers. SE-TDDT differs by connecting these dimensions through climate simulation, growth simulation, MPC, biological guidance, and structured memory.

\begin{landscape}
\begin{table}[p] 
\centering 
\small 
\caption{Coverage-based comparison of SE-TDDT with existing study groups.} 
\label{tab:coverage_comparison} 
\setlength{\tabcolsep}{3pt}
\begin{tabularx}{\linewidth}{@{}P{3.2cm} Y P{1.7cm} P{1.55cm} P{1.7cm} P{1.9cm} P{1.8cm} Y@{}} 
\toprule
Study group & Main focus & Climate / energy & Growth / feed & FE / HP / limitation & Biological feedback to MPC & Single-episode memory & Difference from SE-TDDT \\ 
\midrule
Agricultural and greenhouse DTs~\cite{subeesh2025agricultural, escriba2024digital, qu2023digital} & Monitoring, prediction, environmental control & Yes & Usually no & No & No & No & Limited livestock-growth loop and no biological guidance. \\ 
HVAC, MPC, and Bayesian optimization~\cite{drgona2020mpc, yang2019adaptive, xu2024datadriven, bring1999models, tanaskovic2017robust, fiducioso2019safe, xin2024review, maddalena2022experimental, lin2023bayesian, hosamo2023hvacdt} & Comfort, uncertainty, and energy optimization & Yes & No & No & No & No & Strong climate control, but no growth--feed feedback. \\ 
Livestock-building climate and THI studies~\cite{shin2024thi, costantino2023livestock} & Ventilation, thermal comfort, and barn climate & Yes & Limited & Limited & No & No & Animal-housing context exists, but growth outputs are not returned to MPC. \\ 
Mechanistic beef-growth models~\cite{vanderlinden2019ligaps1, vanderlinden2019ligaps2, vanderlinden2019ligaps3} & Growth, feed intake, digestion, energy/protein use & No & Yes & Yes & No & No & Strong biological model, but not an actuator controller. \\ 
Full SE-TDDT & Climate--growth--feed--energy control with memory & Yes & Yes & Yes & Yes & Yes & Connects climate, growth, biological guidance, MPC, and KStore memory. \\ 
\bottomrule
\end{tabularx}
\end{table}
\end{landscape}

Compared with classical MPC and adaptive building-control approaches, SE-TDDT does not treat barn climate only as a temperature, ventilation, comfort, and energy-control problem. Conventional MPC and Bayesian optimization methods mainly focus on temperature, humidity, ventilation, comfort constraints, uncertainty, and energy consumption~\cite{drgona2020mpc, yang2019adaptive, xu2024datadriven, bring1999models,tanaskovic2017robust, fiducioso2019safe, xin2024review, maddalena2022experimental, lin2023bayesian}. In contrast, SE-TDDT adds biological guidance to the MPC decision process. The growth loop returns heat production, feed intake, feed efficiency, growth status, and limiting factors to the controller, so actuator candidates are evaluated not only by comfort and energy cost, but also by their biological implications~\cite{vanderlinden2019ligaps1, vanderlinden2019ligaps2, vanderlinden2019ligaps3}.

Compared with standalone LiGAPS-Beef-type growth simulation, SE-TDDT connects the biological model to climate control. LiGAPS-Beef can simulate growth, feed intake, digestion, energy/protein use, heat production, average daily gain, and limiting factors at the animal and herd levels~\cite{vanderlinden2019ligaps1, vanderlinden2019ligaps2, vanderlinden2019ligaps3}. However, in its standalone use, these outputs are not directly returned to a real-time climate-control loop. SE-TDDT uses this growth layer as a feedback source: daily aggregated climate data are sent to the growth simulator, and the resulting biological outputs are converted into guidance for MPC.

Compared with general agricultural DTs, SE-TDDT provides a more executable, multi-rate, and edge-aware architecture. Existing agricultural DTs commonly support monitoring, simulation, prediction, and management decisions, but many remain domain-oriented and weakly coupled across climate, growth, feed, and energy~\cite{amiri2025jumeaux, subeesh2025agricultural, escriba2024digital, gonzalez2022monitoring}. SE-TDDT addresses this limitation by using a five-minute inner climate loop, a daily outer growth loop, structured memory, and online edge deployment. Therefore, its main difference is not only the use of a digital-twin concept, but the implementation of a feedback-based architecture in which climate, biological response, energy use, feed-related guidance, and control memory are coordinated within one executable cycle~\cite{amiri2025jumeaux}.

\subsection{Scientific and Implementation Implications}

The main scientific implication of SE-TDDT is that barn climate and livestock growth are treated as feedback-coupled processes rather than as independent subsystems. Climate decisions affect comfort, feed response, heat production, and growth, while the biological state of the herd changes future thermal load, ventilation demand, and energy requirements. Therefore, the proposed framework extends climate control from a single-domain HVAC task to a trans-domain decision process in which growth outputs return to MPC as biological guidance, as shown in Figure~\ref{fig:image_1}.

From an implementation perspective, the edge deployment shows a practical path toward farm-level control. The SE-TDDT workflow was executed on an Orange Pi 5 with CPU-based computing, and the edge controller can load the final knowledge package to perform local monitoring, decision support, and lightweight adaptation without GPU or cloud-side computation. This demonstrates that the framework is not only a simulation concept, but can be organized for low-power barn-level execution under edge-computing constraints, as illustrated in Figure~\ref{fig:image_3}.

The KStore design also shows that single-episode learning can be transformed into reusable operational memory. SS-KStore stores high-frequency training traces and daily summaries in a streaming form, while SETD-KStore compresses the learned episode into a loadable package containing policy snapshots, climate contexts, biological-guidance–decision relationships, reward indicators, coverage, uncertainty, and training summary. As a result, online control does not start from scratch; it starts from structured trans-domain knowledge and is intended to fine-tune decisions with real barn data during future field deployment.

\section{Limitations and Future Work}
\subsection{Observed Limitations}

The current SE-TDDT version demonstrates executable climate–growth–knowledge coupling, but several limitations remain. First, feed pressure is the main biological weakness. Although comfort and growth were preserved and energy behavior was acceptable, feed performance remained weaker than the other dimensions, as reflected by the low feed-consumption accuracy and the feed–growth radar analysis in Figure~\ref{fig:integrated_results_e} and Table~\ref{tab:sarg_phase_summary}. This indicates that SARG-SEL requires stronger phase–diet adaptation and more direct weighting of feed intake, feed efficiency, average daily gain, feed per kilogram of gain, and growth-limiting factors.

Second, actuator behavior was not fully smooth. Even though the safety filter prevented heating–ventilation conflict, the reports still showed frequent command changes, abrupt changes, and reversal oscillations. This limitation means that future MPC tuning should include stronger switching penalties, ramp constraints, minimum dwell time, hysteresis, and smoother weight adjustment to improve practical actuator operation.

Third, the current validation is still limited in scope. The reported results are based on single-episode and simulation-based execution; therefore, broader field validation with real barn data, numerical baseline execution, real ablation tests, and factorial season-by-growth scenarios are still required.

Fourth, the climate model remains a lightweight single-zone representation. This is useful for edge feasibility, but it cannot fully represent spatial heterogeneity in barn geometry, actuator placement, airflow, temperature, and humidity. A lightweight fast fluid dynamics (FFD) or three-dimensional low-order model is therefore required for future versions.

Fifth, barn-generated gas dynamics are not yet modeled. The current framework lacks a meso-scale gas layer for carbon dioxide $\mathrm{CO}_{2}$, methane $\mathrm{CH}_{4}$, ammonia $\mathrm{NH}_{3}$, nitrous oxide $\mathrm{N}_{2}\mathrm{O}$, and water vapor. Such a layer should connect growth-related manure, nitrogen, moisture, and volatile-solid reservoirs to MPC decisions without adding computationally heavy gas models.

\subsection{Future Work}

Future development should strengthen SE-TDDT in five directions. First, the current lightweight single-zone climate model should be upgraded to an edge-executable FFD model or a three-dimensional low-order model. This improvement is needed to represent barn geometry, actuator placement, airflow heterogeneity, temperature distribution, and humidity distribution more accurately than the current single-zone model, while avoiding the high computational cost of full computational fluid dynamics (CFD) and the data requirements of Lattice Boltzmann Method (LBM).

Second, SARG-SEL should be strengthened for phase–diet adaptation. More accurate stage-wise references should be defined for diet phases, and the biological guidance should assign more direct weights to feed intake, feed efficiency, average daily gain, feed per kilogram of gain, and growth-limiting factors.

Third, MPC should include softer actuator constraints to reduce switching and improve command smoothness. Future versions should increase switching penalties, add ramp-rate constraints, define minimum dwell time, use hysteresis, and apply smoother adjustment of decision weights.

Fourth, the framework should be validated online with real barn data. The current results demonstrate single-episode and simulation-based feasibility, but broader validation requires real farm measurements, numerical baseline execution, real ablation tests, and factorial season-by-growth scenarios.

Fifth, a meso-scale gas layer should be added between the growth loop and the climate loop. This layer should model daily gas-production capacity through lightweight reservoirs of nitrogen, manure, moisture, and volatile solids, and incorporate $\mathrm{CO}_{2}$, $\mathrm{CH}_{4}$, $\mathrm{NH}_{3}$, $\mathrm{N}_{2}\mathrm{O}$, and water vapor into MPC decisions without adding computationally heavy gas models.

\section{Conclusion}

This paper presented an SE-TDDT architecture for bio-aware climate and energy control in a closed cattle-fattening barn. The main result shows that climate, growth, energy, feed, biological guidance, and knowledge memory can be kept within a unified and executable decision-making cycle. Therefore, livestock-building climate control can be transformed from a single-domain HVAC problem into a bio-aware and trans-domain control problem.

The scientific contribution of this work is the feedback coupling between climate and growth. Climate decisions affect growth, feed response, and heat production, while growth and heat production also change the future climatic load of the barn. This two-way coupling allows the growth loop to return biological guidance to the MPC layer instead of remaining only a reporting component.

The implementation contribution is the use of single-episode training with SS-KStore and SETD-KStore to transfer offline experience to online edge control. SS-KStore stores high-frequency training traces and daily summaries, while SETD-KStore compresses the learned experience into a loadable knowledge package for online operation. As a result, the edge controller does not start from scratch; it loads prior trans-domain knowledge and is intended to fine-tune decisions with real barn data during future field deployment.

Overall, SE-TDDT demonstrates that barn-level climate control can move beyond conventional HVAC regulation toward an executable, memory-supported, and biologically guided digital-twin architecture.

\FloatBarrier
\section*{Code and Data Availability}

The source code, configuration files, generated datasets, simulation paths, and plotting scripts associated with this study have been published in a public, versioned archive on Zenodo. This archive includes the SE-TDDT workflow, climate and growth simulation interfaces, MPC decision logic, CCLL-SEL and SARG-SEL build scripts, KStore generation scripts, climate datasets, sample inputs, processed outputs, and related scripts. The permanent record is available at: \href{https://doi.org/10.5281/zenodo.21877123}{https://doi.org/10.5281/zenodo.21877123}.

\section*{Ethics Statement}

This study is simulation-based and does not report any new intervention on live animals or any invasive animal experiment. Future online validation using real barn data will comply with the relevant requirements for animal welfare, farm operations, data management, and institutional approval.

\section*{Declaration of Competing Interest}

The author declares no competing financial interests or personal relationships that could have influenced this work.

\section*{Funding}

This research did not receive any specific grant from funding agencies in the public, commercial, or not-for-profit sectors.

\section*{Acknowledgments}

The author acknowledges the published building-climate, livestock-growth, and digital-twin literature used to formulate the SE-TDDT framework, and the processed climate and simulation reports used to construct the CCLL-SEL and SARG-SEL summaries.

\section*{Contributor Roles Taxonomy (CRediT) Authorship Contribution Statement}

Mansoorali Amiri: Conceptualization, Methodology, Software, Formal analysis, Validation, Data curation, Visualization, Writing -- original draft, Writing -- review and editing.

\FloatBarrier
\bibliographystyle{unsrtnat}  
\bibliography{references}

\end{document}